\documentclass[numberedappendix,iop]{emulateapj}
\slugcomment{{\sc accepted for publication in The Astrophysical Journal:} May 24, 2013}

\usepackage{apjfonts}

\begin{document}

\title{Rotating globular clusters}

\author{P. Bianchini\altaffilmark{1}, A.~L. Varri\altaffilmark{2}, G. Bertin, and A. Zocchi} 
\affil{Universit\`{a} degli Studi di Milano, Dipartimento di Fisica, via Celoria 16, I-20133 Milano, Italy; bianchini@mpia.de} 
	

\altaffiltext{1}{now at Max Planck Institute for Astronomy, K\"{o}nigstuhl 17, 69117 Heidelberg, Germany}
\altaffiltext{2}{now at Indiana University, Department of Astronomy, 727 East 3rd Street, Swain West 319, Bloomington, IN 47405-7105, USA}


\begin{abstract}
Internal rotation is considered to play a major role in 
the dynamics of some globular clusters. However, in only few cases it has been studied by 
quantitative application of realistic and physically justified global models.
Here we present a dynamical analysis of the photometry and three-dimensional kinematics of $\omega$ Cen, 47 Tuc, and M15, by means of a recently introduced family of self-consistent axisymmetric rotating models. The three clusters, characterized by different relaxation conditions, show evidence of differential rotation and deviations from sphericity.
The combination of line-of-sight velocities and proper motions allows us to determine their internal dynamics, predict their morphology, and estimate their dynamical distance.
The well-relaxed cluster 47 Tuc is very well interpreted by our 
model; internal rotation is found to explain the observed morphology. For M15, we provide a global model in good agreement with the data, including the central behavior of the rotation profile and the shape of the ellipticity profile. For the partially relaxed cluster 
$\omega$ Cen, the selected model reproduces the complex three-dimensional kinematics; in particular the observed anisotropy profile, 
characterized by a transition from isotropy, to weakly-radial anisotropy, and then to tangential anisotropy in the outer parts. The discrepancy found for the steep central gradient in the observed line-of-sight velocity dispersion profile and for 
the ellipticity profile is ascribed to the condition of only partial relaxation of this 
cluster and the interplay between rotation and radial anisotropy.
\end{abstract}

\keywords{globular clusters:general - globular clusters:individual: NGC 104 (47 Tuc), NGC 5139 ($\omega$~Cen), NGC 7078 (M15)}

\section{Introduction}
Globular clusters (GCs) have long been considered simple quasi-relaxed nonrotating stellar systems, characterized by spherical symmetry and isotropy in velocity space. Spherical isotropic models (in particular, the \citealp{King1966} models and a spherical, nonrotating version of the  \citealp{Wilson1975} models) have indeed been shown to provide a satisfactory zeroth-order description of the main observed dynamical properties (for a recent dynamical study of a large sample of GCs based on the modeling of only the observed photometric profiles, see \citealp{MLvdM2005}, hereafter denoted as MLvdM05; for a dynamical study of a sample of 13 GCs based on both photometric and (line-of-sight) kinematic profiles, see \citealp{Zocchi2012}, hereafter  ZBV12).

However, the acquisition of high-quality data is rapidly bringing us well beyond such simple picture. In particular, deviations from sphericity have been observed (see \citealp{Geyer1983}, \citealp{WhiteShawl1987}, and \citealp{ChenChen2010}; the last two papers will be denoted as WS87 and CC10, respectively). In addition, significant internal rotation has been detected in a growing number of Galactic GCs from line-of-sight velocity measurements (for a summary, see Table 7.2 in \citealp{MeylanHeggie1997}; for more recent investigations, see, among others, \citealp{Lane2011, Bellazzini2012}) and, in a few cases, from kinematical measurements in the plane of the sky (e.g., for M22 see \citealp{Peterson1994}, for $\omega$ Cen see \citealp{LePoole2002}, and for 47 Tuc see \citealp{Anderson2003}). Detailed three-dimensional kinematics are therefore available for selected Galactic clusters. As to the measurement of proper motions, the Hubble Space Telescope (HST) is best used to probe the central regions of the systems \citep{McLaughlin2006,Anderson2010}, whereas ground-based observations are considered for wide-field coverage \citep{vanLeeuwen2000,Bellini2009,Sariya2012,Zloczewski2012}. The future mission GAIA is planned to provide three-dimensional kinematical data for a large number of stars in globular clusters (except for the center of very dense objects affected by crowding). All this progress calls for the development of a more complete and realistic dynamical modeling framework, in which internal rotation and deviations form sphericity are fully taken into consideration.

Internal rotation, external tides, and pressure anisotropy are the main physical factors that could be responsible for the observed flattening of globular clusters, but we still do not know which is the dominant cause of the observed deviations from spherical symmetry (\citealp{vandenBergh2008}). In this paper we will not address the effect of tides because they are expected to act mainly in the outer parts of these stellar systems, in regions outside the focus of the present investigation. The suggestion that internal rotation plays a role in determining the structure and morphology of GCs is not new \citep{King1961,Fall1985}. A tool commonly used to determine the importance of rotation in shaping a stellar system is the $V/\sigma$ vs. $\varepsilon$ diagram \citep{Davies1983, Binney2005, Cappellari2007}. Given their small ellipticities ($\varepsilon < 0.35$), globular clusters are typically located in the portion of the diagram representing configurations characterized by solid-body rotation and isotropy to mild anisotropy of the velocity dispersion tensor. However, this approach provides only a zeroth-order description of the dynamical interplay between internal rotation and pressure anisotropy. In particular, such diagram considers only global quantities, which can vary significantly as a result of detailed changes with radius of the anisotropy parameter and of inclination effects, factors that are generally not well constrained observationally (see Sect.~7.3 and Fig.~\ref{fig:v_sigma_graph}). Therefore, the present investigation is  motivated by the need to provide a more realistic dynamical interpretation of selected rotating Galactic globular clusters, with particular attention to objects that show small yet significant deviations from the behavior of a simple isotropic (slow) rotator. In this respect, the most significant investigations made so far are the orbit-based axisymmetric modeling of $\omega$ Cen and M15 (\citealp{vandeVen2006,vandenBosch2006}, respectively), the study of $\omega$ Cen by means of axisymmetric \citealp{Wilson1975} models \citep{Sollima2009} and an oblate rotator nonparametric model \citep{Merritt1997}, the description of M13 by means of a family of models with three integrals of the motion \citep{LuptonGunn1987,Lupton1987}, and the analysis of the internal dynamics of a small sample of Galactic GCs through dedicated 2D Fokker-Planck models \citep{Fiestas2006}.

Additional interest in the role of rotation derives from the fact that the presence of global angular momentum is expected to change the long-term dynamical evolution of stellar systems with respect to the traditional paradigm based on nonrotating models (for a summary, see \citealp{HeggieHut2003}). Numerical investigations, primarily based on a Fokker-Planck approach \citep{EinselSpurzem1999,Kim2002,Kim2008,Hong2012}, demonstrate that, in general, the presence of rotation accelerates dynamical evolution.
 
Internal rotation may also play an indirect role in the controversial issue of the presence of Intermediate Mass Black Holes (IMBH) in GCs. In fact, sizable central gradients in the velocity dispersion profiles are often ascribed to the presence of an IMBH \citep{Baumgardt2005}. A critical discussion of the observed gradients is often reduced to the application of the Jeans equations in which variations of the slope of the velocity dispersion profile are obtained by varying only the amount of pressure anisotropy (without considering rotation; e.g., see \citealp{Lanzoni2013}, \citealp{Lut11}, and \citealp{vdMA2010}). However, differential rotation and pressure anisotropy can cooperate to produce nontrivial gradients in the velocity dispersion profiles (see \citealp{Varri2012}, hereafter denoted as VB12) and might thus be an important element to be considered in the interpretation of the data. 

In view of these motivations, a new family of self-consistent axisymmetric models has been introduced recently, specifically designed to describe quasi-relaxed stellar systems with finite global angular momentum (VB12); the models are characterized by differential rotation, approximately rigid in the center and vanishing in the outer parts, and pressure anisotropy.

In the present paper we apply this family of differentially rotating global models to three Galactic GCs, namely $\omega$ Cen, 47~Tuc, and M15, that have been observed in detail and are known to exhibit evidence for rotation. In Sect.~\ref{observations} we present the available data sets for these three GCs and describe the procedure followed to construct the profiles of the relevant photometric and kinematic quantities (some important detailed description is provided separately in Appendices A and B). In Sect.~\ref{selection} we summarize the property of the adopted family of self-consistent rotating dynamical models and introduce the method used to identify the best model to describe the data available for the three clusters. The detailed results on $\omega$~ Cen, 47 Tuc, and M15 are reported in Sects.~\ref{omega}, \ref{47}, and~\ref{M}, taking into consideration that $\omega$~Cen is only partially relaxed while 47 Tuc and M15 are fully relaxed. In Sect. \ref{comparison} we discuss the results of the present paper and compare them with those obtained from previous studies. Finally, in Sect.~\ref{conclusions}  we summarize the conclusions that can be drawn from our study.

\label{sec:model}
\begin{table*}
\begin{center}
\caption[Properties of $\omega$ Cen, 47 Tuc, and M15.]{Properties of $\omega$ Cen, 47 Tuc, and M15.} \label{tab:1}
\begin{tabular}{lccccccccccc}
\hline\hline
GC &d&$R_{\mathrm{c}}$&C&$\log T_{\mathrm{c}}$&\multicolumn{2}{c}{$\varepsilon$} &$\phi$&PA& i  &$N_{\mathrm{los}}$&$N_{\mathrm{pm}}$\\
&(1)&(2)&(3)&(4)&(5)&(6)&(7)&(8)&(9)&(10)&(11)\\
\hline
$\omega$ Cen&$5.2\pm0.7$&$142.20\pm8.26$&$1.31\pm0.04$&$9.52\pm0.04$&$0.21\pm0.02$&$0.17\pm0.00$&$6\pm0$&$12\pm1$&$50\pm4$&1868&2740 + 72\,970\\
47 Tuc&$4.5\pm0.2$&$21.60\pm1.31$&$2.07\pm0.03$&$7.85\pm0.07$&$0.16\pm0.02$&$0.09\pm0.01$&$123\pm1$&$136\pm1$&$\approx45$&2476&12\,974\\
M15&$10.4\pm0.8$&$8.40\pm0.95$&$2.29\pm0.18$&$7.62\pm0.06$&$0.19\pm0.10$&$0.05\pm0.00$&$215\pm1$&$106\pm1$&$60\pm15$&1777&703\\
\hline
\end{tabular}
\tablecomments{For each cluster we list: (1) the distance from the Sun d in kpc; (2) the core radius $R_{\mathrm{c}}$ in arcsec; (3) the concentration parameter $C$; (4) the logarithm of the core relaxation time $T_{\mathrm{c}}$ in years from spherical King models; the ellipticity $\varepsilon=1-b_p/a_p$ (where $a_p$ and $b_p$ indicate the observed major and minor axes), as reported by (5) CC10  and (6) WS87; (7) the position angle of the photometric minor axis $\phi$ measured in degrees (East of North); (8) the position angle of the kinematic rotation axis PA on the plane of the sky measured in degrees (East of North); (9) the inclination $i$ of the rotation axis with respect to the line-of-sight measured in degrees; the number of data points for the samples of (10) line-of-sight velocities $N_{\mathrm{los}}$ and (11) proper motions  $N_{\mathrm{pm}}$.}
\tablerefs{From Col. (1) to Col. (3), \citet{HarrisCat2010}; Col. (4) ZBV12; Col. (5) CC10; Cols. (6) and (7) WS87; Col. (9) \citet{vandeVen2006}, \citet{Anderson2003}, \citet{vandenBosch2006} (from top to bottom, that is, for $\omega$ Cen, 47 Tuc, and M15, respectively); Cols. (8), (10), and (11) considered in the present work. }
\end{center}
\end{table*}
\section{Observed kinematic and photometric profiles}
\label{observations}

In this section we describe the kinematic and photometric data sets that we will use in the dynamical analysis and the methods to build the relevant profiles, with particular attention to the construction of the rotation profiles. Table \ref{tab:1} summarizes the basic properties of the globular clusters $\omega$ Cen, 47 Tuc, and M15. 

\subsection{Kinematic profiles}
\label{kinprof}
We gathered and combined kinematic data sets taken from the literature to cover a large radial extent. In Appendix \ref{App.A} we describe in detail the data sets selected for the line-of-sight velocities and proper motions. The data are referred to a Cartesian coordinate system $(x_p,y_p)$, with $x_p$ and $y_p$ aligned with the major and minor axes, respectively \citep{vandeVen2006}. The $z_p$ axis identifies the line-of-sight direction. Proper motions are then decomposed into projected tangential $\mu_t$ and radial $\mu_R$ components.

The present dynamical study is based on a combined analysis of the following kinematic profiles: (1) rotation profiles, (2) velocity dispersion profiles, and (3) the pressure anisotropy profile. The kinematic profiles are constructed with the traditional binning approach, that is, the data are divided into bins containing an equal number of stars. In particular, radial bins are used to construct the velocity dispersion and anisotropy profiles, whereas the line-of-sight rotation profile is constructed by binning along the observed major axis, in intervals of $x_p$. We choose a number of bins that represents the best compromise between having a rich radial sampling and accurate points,\footnote{The number of stars per bin is chosen to be large enough to limit the uncertainties associated with low-number statistics (for the profiles constructed in this paper the number of data per bin is $> 90$).}as in ZBV12.

To calculate the mean velocity and the velocity dispersion, with the associated errors, we apply a Maximum Likelihood technique to the data, following the method described by Pryor \& Meylan (1993) in which non-constant velocity errors are taken into consideration.\footnote{A contamination model is not included in the Maximum Likelihood estimator, since potential non-members have already been excluded in the data sets that have been considered.} The details of the procedure used to obtain the different profiles are given below.

\subsubsection{Rotation profiles}
\label{rotation}
\begin{figure}[t]
\centering
\includegraphics[width=0.40\textwidth]{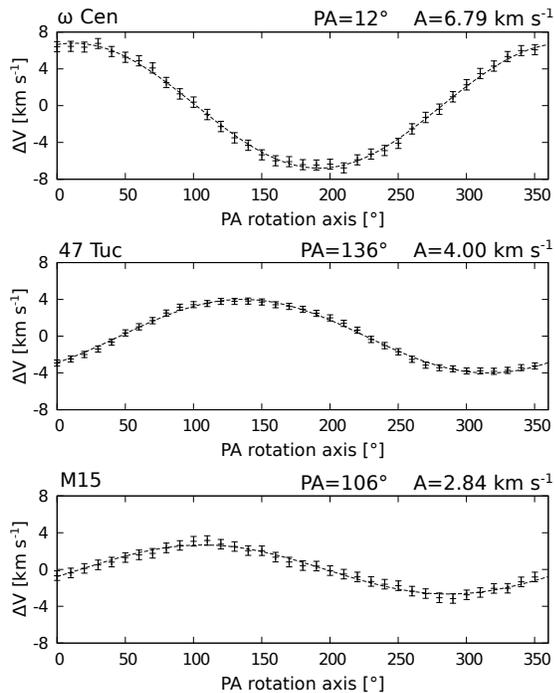}
\caption[Difference of the mean velocities]{Difference of the mean velocities calculated on each side of the system divided by a line passing through the center with a given position angle PA. The PA at which the maximum difference is reached corresponds to the position of the rotation axis. The best-fit sine function is plotted (solid line) and the corresponding PA and amplitude $A$ are indicated.}
\label{fig:PA}
\end{figure}

The first step in building a rotation profile consists in identifying the position angle (PA) of the projected rotation axis in the plane of the sky (defined as the angle between the rotation axis and the North direction, measured East of North). To identify the PA the following standard procedure is used (e.g., see \citealp{Cote1995,Bellazzini2012}): the line-of-sight velocities data set is divided in two halves by a line passing through the center with a given PA and for each subsample the mean line-of-sight velocity is computed; the PA is varied in steps of $10^\circ$ and the difference between the mean velocities $\Delta V$ is plotted against PA. The resulting pattern is fitted with a sine function (see Fig.~\ref{fig:PA}): the PA at which the maximum difference in mean velocities is reached corresponds to the rotation axis and the amplitude $A$ of the sine function gives an estimate of the significance of the internal rotation. The values obtained for the PA are used to rotate the Cartesian coordinate system in the plane of the sky by aligning $x_p$ and $y_p$ with the major and minor axes, respectively (Appendix \ref{App.A}). The results are listed in Table~\ref{tab:1} and compared to the position angles of the photometric minor axes $\phi$ reported by WS87. 

The position angles of the kinematic minor axes of $\omega$ Cen and 47 Tuc are in reasonable agreement with the photometric ones, suggesting a direct connection between the presence of internal rotation and observed flattening. A discrepancy is found instead for M15: for this cluster the small observed flattening ($\varepsilon\approx0.05$) makes the identification of the minor axis nontrivial. Various estimates of the photometric position angle are given in the literature, ranging from $215\degr$ to $135\degr$, suggesting a possible twisting of the position angle of both the photometric and kinematic minor axes \citep{Gebhardt2000,vandenBosch2006}. Additional tests on the twisting of the rotation axis and on the radial variation of the rotation amplitude are recorded in Appendix \ref{App.B}.

After identifying the rotation axis, we can proceed to build the rotation profiles. First we subtract from each data set the measured mean systemic velocity; then we divide the line-of-sight velocities data set in bins along the major axis $x_p$; each bin is assigned an average $x$ position, mean velocity, and associated uncertainty. In the case of the proper motion data set, the rotation profile is constructed by dividing the data set in radial bins and by computing for each of them the mean radial distance and the mean velocity, separately for the tangential and projected radial components. We then end up with three mean-velocity profiles, one for the line-of-sight, $V_\mathrm{los}(x_p)$, and two for the proper motions, $V_\mathrm{t}(R )$ and $V_\mathrm{R}(R )$.

\subsubsection{Velocity dispersion and anisotropy profiles}

The velocity dispersion profiles are computed by dividing the data sets into radial bins; we consider the mean velocity of the entire data set as a constant value throughout the cluster, and we calculate the velocity dispersion for each bin with the associated uncertainty. For each bin, the distance from the center is taken to be the mean radial positions of the stars that it contains. The profiles obtained are $\sigma_\mathrm{los}(R )$, $\sigma_\mathrm{t}(R )$, and $\sigma_\mathrm{R}(R )$, respectively for the line-of-sight velocities, projected tangential component of proper motions, and projected radial component of proper motions.

From the dispersion profiles of the proper motions we also calculate the anisotropy profile: this is defined here as the ratio of the velocity dispersion in the tangential component to the velocity dispersion in the radial component, $\sigma_\mathrm{t}(R )/\sigma_\mathrm{R}(R )$. Values of $\sigma_\mathrm{t}/\sigma_\mathrm{R}\approx1$ indicate isotropy in velocity space, $\sigma_\mathrm{t}/\sigma_\mathrm{R}>1$ indicate the presence of tangential anisotropy, and $\sigma_\mathrm{t}/\sigma_\mathrm{R}<1$ radial anisotropy.

\subsection{Photometric profiles}
\label{photprof}
The photometric quantities that we will use in the dynamical analysis are the surface brightness profile and the ellipticity profile. Below we briefly describe the data sets available for the construction of these profiles. 

\subsubsection{Surface brightness profiles}
The surface brightness profiles are taken from ZBV2012: they are V-band surface brightness profiles, built from the data of \citet{TKD1995} divided into circular annuli, so that the surface brightness measured in mag~arcsec$^{-2}$ is reported as a function of projected radius. The profiles are extinction corrected, under the assumption of constant extinction over the entire extent of the cluster.

Since the central regions correspond to the least reliable parts of the profiles of \citet{TKD1995}, a combination of different data sets is needed. The more accurate data available from \citet{Noyola2006} are used for 47 Tuc and M15. For 47 Tuc the data from the two sources are simply co-added; for M15 the two data sets are combined by removing the points from \cite{TKD1995} that do not agree with the more recent profile. In the case of $\omega$ Cen the inner points kindly provided by Eva Noyola \citep{Noyola2008} are added to the \citet{TKD1995} surface brightness profile.

\subsubsection{Ellipticity}

From the morphological point of view, globular clusters present only small deviations from spherical symmetry. Yet, there is observational evidence of flattening, as measured by the ellipticity parameter, defined as $\varepsilon= 1-b_p/a_p$, where $b_p/a_p$ is the ratio of the minor to major axis of the projected image of a cluster in the plane of the sky. For a long time, the WS87 database represented the only comprehensive collection of ellipticity measurements for the Galactic globular clusters; recently, an alternative homogeneous database of ellipticities has been published by CC10. The two distributions of values show significant differences: in fact, from the WS87 database (93 objects), Galactic clusters appear to be predominantly round, with the peak of the distribution at $\varepsilon \approx 0.05$, maximum value at $\varepsilon \approx 0.3$, and axial ratios randomly oriented in space. In contrast, the distribution of the CC10 ellipticities (116 objects, 82 in common with the other database) is peaked at $\varepsilon \approx 0.15$ and with a maximum value of $\varepsilon \approx 0.45$. In addition, especially for the clusters in the region of the Galactic bulge, their major axes point preferentially toward the Galactic center.

The apparent discrepancies between the two studies should be interpreted by taking into account that (i) WS87 ellipticities result from an optical study, with the use of a surface photometry technique based on the blurring of the digitized images of blue sensitive photographic plates. In turn, CC10 ellipticities are determined with a number-count technique, based on the analysis of the spatial distribution of 2MASS point sources; (ii) as a result of the different resolution limits of the two approaches, WS87 measurements mostly refer to the inner regions of the clusters, whereas CC10 measurements refer to the outer parts. Unfortunately, in both cases, the flattening values do not refer to a standard isophote, such as the cluster half-light radius \citep[see][]{Kontizas1989}. This is an intrinsic limitation, because there is observational evidence that the ellipticity of a cluster depends on radius \citep[see][]{Geyer1983}.
 
In the present paper we will use the ellipticity profile of $\omega$~Cen taken from \citet{Geyer1983}. It is the most extended ellipticity profile available for a Galactic globular cluster, as it reaches $\approx0.5\,r_\mathrm{tr}$, where $r_\mathrm{tr}$ represents the standard truncation radius. In addition, \citet{Anderson2010} report the ellipticity profile of the central region ($R\la250$ arcsec); in the following analysis both data sets will be taken into consideration. For 47 Tuc and M15 we will use the profiles of Fig.~5 in WS87. They reach $\approx0.2\,r_\mathrm{tr}$ and $\approx0.4\,r_\mathrm{tr}$, respectively. We note that a genuine radial variation is present in the three ellipticity profiles. This is particularly evident for $\omega$~Cen, which exhibits a nonmonotonic behavior.

\section{Model identification and predictions}
\label{selection}

\begin{table*}[t!]
\caption{Kinematic quantities used to identify the dimensionless parameters of a model.}
\label{tab:3}
\centering
\begin{tabular}{lcccccccc}
\hline\hline
GC & $\sigma_0$ & $V^{\mathrm{rot}}_{\mathrm{max}}$ & $V^{\mathrm{rot}}_{\mathrm{max}}/\sigma_0$ & $R_\mathrm{h}$ & $R^{\mathrm{rot}}_{\mathrm{max}}$&$R^{\mathrm{rot}}_{\mathrm{max}}/R_\mathrm{h}$& $R_\mathrm{a}$& $R_\mathrm{a}/R_\mathrm{h}$\\
			&(1)		&(2)	        		  &(3)		      &(4)&(5)&(6)&(7)&(8)\\
\hline
$\omega$ Cen &$17.31\pm1.72$ 	& $5.80\pm0.32 $		  & $0.34\pm0.04$ &   $300.06\pm3.51$  &$510.10\pm10.21$			&$1.69\pm0.04$ & 1035.21$\pm32.10$   &$3.45\pm0.11$\\
47 Tuc 		&$13.06\pm1.00$	& $3.26\pm 0.40$		  & $0.25\pm0.04$ &   $190.22\pm3.06$  &$342.40\pm5.13$ 		& $1.80\pm0.04$ &\ldots &\ldots\\
M15		 	&$12.93\pm1.06$ 	& $3.00\pm0.63$		  & $0.23\pm0.05$ &   $60.26\pm9.70$ & $79.34\pm12.54$& 		$1.32\pm0.30$& \ldots&\ldots\\
\hline
\end{tabular}
\tablecomments{For each cluster we report in Col.~(1) the observed central line-of-sight velocity dispersion $\sigma_0$ in km s$^{-1}$, in Col.~(2) the maximum of the line-of-sight rotation profile $V^{\mathrm{rot}}_{\mathrm{max}}$ in km s$^{-1}$, in Col.~(3) the ratio $V^{\mathrm{rot}}_{\mathrm{max}}/\sigma_0$, in Col.~(4) the half-light radius $R_\mathrm{h}$ in arcsec from \citet{HarrisCat2010}, in Col.~(5) the position of the maximum of the rotation profile $R^{\mathrm{rot}}_{\mathrm{max}}$ expressed in arcsec,  in Col.~(6) the ratio $R^{\mathrm{rot}}_{\mathrm{max}}/R_\mathrm{h}$ , in Col.~(7) the position $R_\mathrm{a}$ of the transition from the regime of radial anisotropy to tangential anisotropy in arcsec, and in Col.~(8) the ratio $R^{\mathrm{a}}/R_\mathrm{h}$. A blank space in the last two columns indicates that the desired information is not available from the data. Columns (3), (6), and (8) guide our choice of the three dimensionless parameters that characterize the internal structure of the models.}
\end{table*}

The family of self-consistent axisymmetric models that we will consider has been specifically designed to describe quasi-relaxed stellar systems with finite global angular momentum (VB12). These models are global, finite-mass solutions of the self-consistent problem associated with the distribution function $f_{WT}^d(I)$ [see Eq.~(21) in VB12], in which the integral of the motion $I=I(E,J_z)$ is defined as
\begin{equation}
I(E,J_z)=E-\frac{\omega J_z}{1+bJ_z^{2c}},
\end{equation}
where $\omega$, $b$, and $c>1/2$ are positive constants. The subscript $WT$ in the distribution function is a reminder that the form of the function is that of the corresponding spherical isotropic nonrotating models characterized by Wilson truncation; a full description of the physical arguments that have led to this choice of distribution function is provided in VB12. The integral of the motion reduces to $I\sim E$ for high values of $J_z$ and to the Jacobi integral $I\sim H=E-\omega J_z$ for low values of $J_z$. Therefore, the models are characterized by differential rotation, approximately rigid in the center and vanishing in the outer parts. The models are defined by four dimensionless parameters. Two dimensionless parameters are the concentration parameter $\Psi$ (this parameter is often denoted as $W_0$ in the description of the King models), defined as the depth of the dimensionless potential well at the center of the cluster, and the rotation-strength parameter $\chi=\omega^2/(4\pi G\rho_0)$. The parameters $b$ (or, equivalently, the dimensionless parameter $\bar{b}$; see definition in VB12) and $c$ determine the shape of the rotation profile. For the purposes of the present study, we checked that a variation of $c$ does not introduce significant differences, and thus we decided to simplify our investigation by setting $c = 1$. The effect of taking a larger value of $\bar{b}$ is to produce models in which the solid-body rotation covers a wider radial range.
For the self-consistent models, the velocity dispersion tensor is characterized by isotropy in the central region, weak radial anisotropy in the intermediate regions, and tangential anisotropy in the outer parts.\footnote{Tangentially-biased pressure anisotropy in the outer parts of a star cluster is considered to be a natural result of the dynamical evolution of a stellar system within an external tidal field, which induces a preferential loss of stars on radial orbits (this effect has been studied primarily by means of Fokker-Plank and N-body simulations; e.g., see \citealp{Takahashi2000,Baumgardt2003,HurleyShara2012}).}$^{,}$\footnote{Self-consistent models characterized by the presence of tangential anisotropy are rare \citep[see also][]{An2006}; so far, the only dynamical model (of $\omega$ Cen) in which the observed tangential anisotropy has been properly taken into account is the descriptive Schwarzschild model constructed by \citet{vandeVen2006}.} The behavior of the pressure tensor in the external regions of a configuration was not assigned {\em a priori} in the definition of the models: it results from the requirement of self-consistency and from the relevant truncation prescription in phase space.

To carry out  the comparison between our differentially rotating models and the observations, we have to specify three dimensionless parameters (the concentration $\Psi$, the rotation strength $\chi$, and the parameter $\bar{b}$) and five additional quantities. Three physical scales (i.e., the radial scale $r_0$, the central surface brightness $SB_0$, and the velocity scale $v_0$). Finally, the inclination angle $i$ between the rotation axis and the line-of-sight direction, and the distance to the cluster (required to convert the proper motions in km s$^{-1}$). Such a highly-dimensional parameter space is likely to lead to a high degree of degeneracy. Therefore, we decided to separate the modeling procedure in three steps, by starting from the focus of interest of this paper, that is the presence of internal rotation. First, we determine the dimensionless structural parameters by following few natural selection criteria based on the observed kinematics, then we set the physical scales by means of a few standard statistical fits (this information will be summarized in Tables \ref{tab:5} and \ref{tab:6}), and finally we test some properties of the models as predictions in relation to other observational data not used in the first two steps.

The exploration of the complete 3D dimensionless parameter space is guided by the following general properties of the models: (1) large values of the concentration parameter $\Psi$ determine spatially extended configurations, in terms of the relevant units of length (see VB12 for details); (2) configurations characterized by a given value of concentration and increasing values of the rotation strength parameter $\chi$ are progressively more compact because of the adopted truncation prescription in phase space; (3) the parameter $\bar{b}$ determines the shape of the line-of-sight rotation profile, in particular, it regulates the radial position of the velocity peak.

\begin{table*}[t]
\caption[]{Dimensionless parameters and physical scales of the best-fit models.}
\label{tab:5}
\centering
\begin{tabular}{lcccccccccc}
\hline\hline
 &  &\multicolumn{3}{c}{Dimensionless parameters} & & \multicolumn{3}{c}{Physical scales} && Dynamical distance\\
\cline{3-5}  \cline{7-9} \cline{11-11}
GC & & $\Psi$ & $\chi$ & $\bar{b}$ & & $\mathrm{SB}_0$ & $r_0$ & $v_0$ & &$d$ \\
&&(1)&(2)&(3)& &(4)&(5)&(6)& &(7)\\
\hline

$\omega$ Cen  & & 5.8 &$14.4\times10^{-3}$&0.040&  & $16.43\pm0.05$ & $134.54\pm1.13$  & $15.87\pm0.27$ & & $4.11\pm0.07$  \\
47 Tuc  & & 7.6 & $1.6\times10^{-3}$ & 0.008 &  & $14.30\pm0.08$  & $24.41\pm0.14$ & $13.35\pm0.21$ &  &$4.15\pm0.07$ \\
M15  & & 6.8 & $1.6\times10^{-3}$ & 0.035  &  & $14.65\pm0.01$  &  $13.33\pm0.20$  & $12.52\pm0.24$ &  & $10.52\pm0.38$ \\
\hline
\end{tabular}
\tablecomments{For each cluster we list: the concentration parameter $\Psi$ in Col. (1), the rotation strength parameter $\chi$ in Col. (2), the $\bar{b}$ parameter in Col. (3), the V-band central surface brightness $\mathrm{SB}_0$ in mag arcsec$^{-2}$ in Col. (4), the radial scale $r_0$ in arcsec in Col. (5), the velocity scale $v_0$ in km s$^{-1}$ in Col. (6), and the best-fit dynamical distance d in kpc in Col. (7); for the physical scales and the distance, the associated $1\sigma$-errors are also shown. Note that $r_0$ is an intrinsic quantity; it is recorded here in arcseconds, for easier comparison with the observations.}
\end{table*}

\begin{table*}[t]
\caption{Derived parameters.}
\label{tab:6}
\centering
\begin{tabular}{lccccccc}
\hline\hline
GC & $C$ & $R_\mathrm{c}$ & $R_\mathrm{h}$& $r_\mathrm{tr}$ & $M$ & $M/L_\mathrm{V}$& $\log\rho_0$ \\
&(1)&(2)&(3)&(4)&(5)&(6)&(7)\\
\hline
$\omega$ Cen & $1.27\pm0.01$& $127.8\pm1.1$& $282.5\pm2.4$& $2400.3\pm20.2$&$19.53\pm0.16$&$2.86\pm0.14$&$3.737\pm0.034$\\
47 Tuc &$1.87\pm0.01$	&$24.6\pm0.1$ & $162.8\pm0.9$& $1814.9\pm10.4$&$6.23\pm0.04$ &$1.69\pm0.13$&$5.090\pm0.102$\\
M15 & $1.94\pm0.02$& $12.9\pm0.2$& $43.7\pm0.7$& $1118.9\pm16.8$ &$4.55\pm0.07$&$1.45\pm0.05$&$4.752\pm0.130$\\
\hline
\end{tabular}
\tablecomments{For each cluster we provide the structural parameters derived for the best-fit models: (1) the concentration parameter $C=~\log(r_\mathrm{tr}/R_\mathrm{c})$, (2) the projected core radius $R_\mathrm{c}$ in arcsec, (3) the projected half-mass radius $R_\mathrm{h}$ in arcsec, (4) the truncation radius $r_\mathrm{tr}$ in arcsec, (5) the total mass of the cluster $M$ in units of $10^5M_\odot$, (6) the V-band mass-to-light ratio in solar units, (7) the logarithm of the central mass density $\rho_0$ in units of $M_\odot$~ pc$^{-3}$.}
\end{table*}

\subsection{Dimensionless parameters}
\label{dimensionless}
From Sect. \ref{rotation} it is clear that the globular clusters under consideration are characterized by significant global internal rotation. Therefore, we start from the observed rotation properties to identify the natural ranges of the three dimensionless parameters. In particular, the parameters should lead to configurations that successfully reproduce the following observations: (1) the observed value of $V^{\mathrm{rot}}_{\mathrm{max}}/\sigma_0$, that is, the ratio of the peak of the rotation velocity profile to the central velocity dispersion for the line-of-sight kinematic data; (2) the observed shape of the rotation profile along the line-of-sight, in particular the position $R^{\mathrm{rot}}_{\mathrm{max}}$ of the rotation peak (relative to the cluster half-light radius); (3) the qualitative behavior of the anisotropy profile (when available), in particular the radial position $R_\mathrm{a}$ (relative to the half-light radius) of the transition from radial anisotropy to tangential anisotropy. The relevant observational quantities to be matched by application of the above selection criteria are calculated and listed in Table~\ref{tab:3}. Specifically, the central velocity dispersions $\sigma_0$ and associated errors are average values calculated from the kinematic data within $R_\mathrm{c}/2$; the peak of rotation $V^{\mathrm{rot}}_{\mathrm{max}}$, its radial position $R^{\mathrm{rot}}_{\mathrm{max}}$ and the radial position of the transition from radial anisotropy to tangential anisotropy $R_\mathrm{a}$ are calculated by fitting a polynomial to the rotation profile and to the anisotropy profile, in the relevant radial ranges. 

Given a set of parameters ($\Psi,\chi,\bar{b}$), the models are projected on the plane of the sky by assuming a known inclination angle $i$, as reported in Table~\ref{tab:1}. The projection is performed by sampling from the relevant distribution function a discrete set of $N=2\,048\,000$ particles and then by performing a rotation of such discrete system to match the relevant inclination angle. The theoretical kinematic and photometric profiles\footnote{The profiles thus constructed are discrete profiles, which are then interpolated to obtain continuous profiles. The statistical scatter associated with the use of discrete model-points is well under control, given the high number of sampling particles considered.} are then constructed by following the procedures described in Sect. \ref{kinprof} and \ref{photprof}. The central dispersion $\sigma_0$, the maximum of the rotation profile $V^{\mathrm{rot}}_{\mathrm{max}}$, and its position $R^{\mathrm{rot}}_{\mathrm{max}}$ are calculated in view of the above-mentioned selection criteria. As to the morphological aspects, the projected isodensity contours are calculated based on the projected number density distribution.
The relevant ellipticity profiles are then constructed by considering the ratio of the principal axes of approximately one hundred isodensity contours, corresponding to selected values of the normalized projected number density in the range $[0.9,10^{-3}]$; smooth profiles are then obtained by performing an average on subsets made of ten to twenty individual ellipticity values (depending on the concentration of the configuration).      

The dimensionless parameters are varied until the kinematic selection criteria are reasonably met,\footnote{Note that the procedure adopted to determine the values of the dimensionless parameters that characterize the internal structure of the models does not allow us to calculate the related formal errors. In any case we will estimate the range of variation of reasonable models (also in relation to the lack of information on $R_a$ for two of the three clusters) by performing a simple exploration of the available parameter space, as described in Sect. \ref{exploration}.} that is, until we obtain models consistent within the uncertainties with the observed quantities listed in Table \ref{tab:3}.

\subsection{Physical scales}
\label{physical}
Once a set of dimensionless parameters is identified, we proceed to determine the relevant physical scales. This is done by fitting the models to the observed profiles, that is, by minimizing the related chi-squared. Two fits are performed. With the photometric fit to the surface brightness profile we determine two scales: the central surface brightness SB$_0$ and the radial scale $r_0$ [the scale $r_0$ is the standard length scale of King models; e.g., see Eq.~(A.2) in ZBV12]. Once SB$_0$ and $r_0$ have been fixed, the velocity scale $v_0$ is determined by means of the kinematic fit, which is performed by minimizing a combined chi-squared defined as the sum of the contributions from the line-of-sight rotation and the line-of-sight velocity dispersion profiles. Finally, the mass-to-light ratio is directly connected to the central surface brightness by the following relation $M/L_\mathrm{V}=\hat{\Sigma}(0) 10^{\mathrm{SB}_0/2.5}$, where $\hat{\Sigma}(0)$ denotes the central surface density expressed in the relevant units. The details of the fitting procedure and of the calculation of the errors are given in Appendix B of ZBV12.

\subsection{Dynamical distance measurement}
\label{distance}
The kinematic information associated with the proper motions is used to measure the distance to the cluster. The relation between proper motions $\mu$ measured in mas yr$^{-1}$ and proper motions $v$ expressed in km~s$^{-1}$ is  
\begin{equation}
\left[\frac{v}{\mbox{km/s}}\right]=4.74\left[\frac{d}{\mbox{kpc}}\right]\left[\frac{\mu}{\mbox{mas/yr}}\right],
\end{equation}
where $d$ is the distance from the observer to the globular cluster. Therefore, with all the dimensionless parameters and physical scales fixed from the previous analysis, we obtain a best-fit distance $d$ (hereafter referred to as dynamical distance) by a combined fit to the observed tangential $\sigma_t$ and radial $\sigma_R$ velocity dispersion profiles (i.e., by minimizing a combined chi-squared defined as the sum of the contributions of the two velocity dispersion profiles in the plane of the sky).

\subsection{Predicted profiles}
\label{predicted}
At this stage for a given cluster the model and the relevant scales have all been determined. A number of other observable quantities are then predicted and can be compared to the available observations. In particular, we wish to include in this category the following quantities: the anisotropy profile $\sigma_t/\sigma_R$, the proper motion mean-velocity profiles $V_t$ and $V_R$, the ellipticity profile $\varepsilon$, and the 2D structure of the isodensity contours which need not be perfect ellipses.

\subsection{Exploration of the parameter space}
\label{exploration}
 
The procedure adopted for the selection of a rotating model gives priority to the kinematic data, which are usually affected by large uncertainties and often do not cover a sufficiently wide radial extent. Therefore, it is important to check whether the selection procedure might be improperly sensitive to these uncertainties. In order to do so, we perform an exploration of the available dimensionless parameter space (Sect. \ref{dimensionless}) by estimating what range of parameters would be consistent with the uncertainties associated with the kinematic observed quantities listed in Table \ref{tab:3}.
 
\begin{figure*}[t]
\centering
\includegraphics[width=0.80\textwidth]{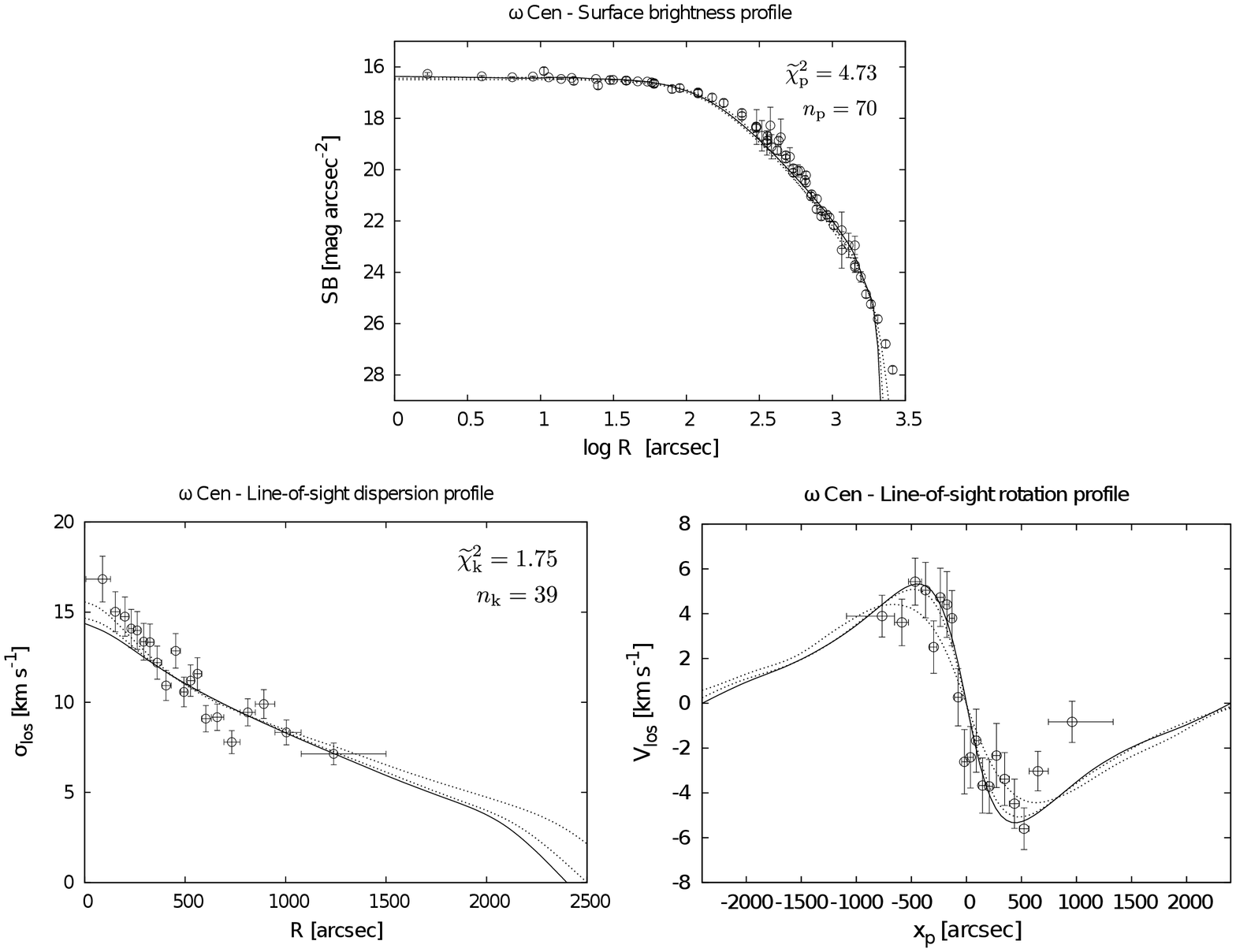}
\caption{Surface brightness profile, line-of-sight velocity dispersion profile, and line-of-sight rotation profile (measured along the projected major axis; for the definition of the $x_p$ coordinate and the way the data are binned, see Sect.~2) for $\omega$ Cen. Solid lines represent the selected model profiles and open circles the observational data points. Vertical bars represent the measured errors and horizontal bars indicate the size of the bins. The fits on these profiles have been used to determine the three physical scales of the model $(\mathrm{SB}_0,r_0,v_0)$ (see Table \ref{tab:5}); the associated photometric and kinematic reduced chi-squared and the number of degrees of freedom are shown (we recall that the kinematic fit is performed by minimizing a combined chi-squared that includes the contributions of both the line-of-sight velocity dispersion profile and the rotation profile, see Sect. \ref{physical}). The light (dotted) lines represent the profiles of the models used to test the sensitivity of the selection procedure to the specific choice of kinematical parameters on which the procedure in based, as described in Sect. \ref{exploration}.}
\label{fig:fit_omega}
\vspace{0.5cm}
\includegraphics[width=0.80\textwidth]{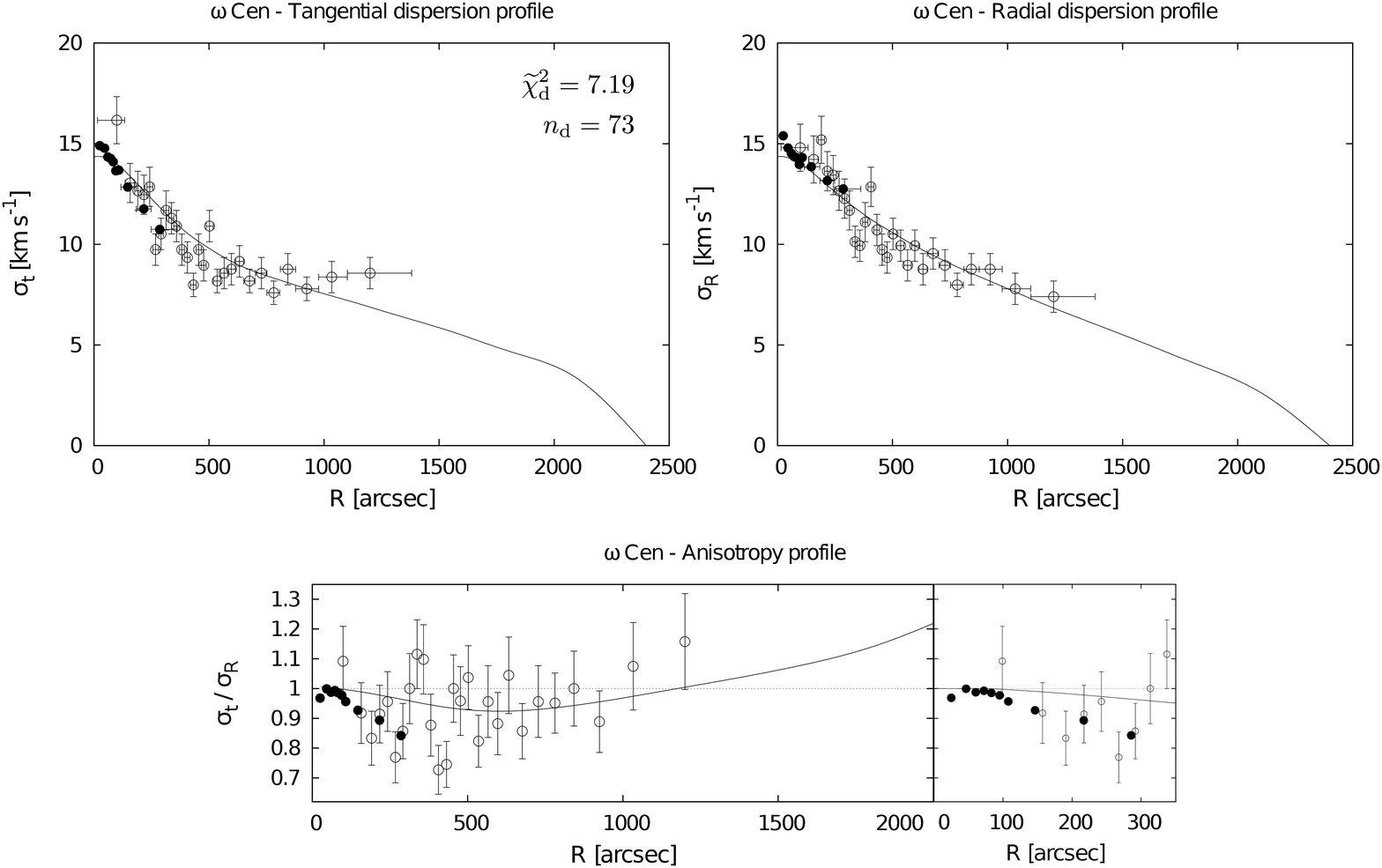}
\caption{The top panels illustrate the fit to the proper motion dispersion profiles along the projected tangential and radial directions for $\omega$ Cen; this fit has determined the dynamical distance $d$. The associated reduced chi-squared and the number of degrees of freedom are shown. The bottom panels show the predicted anisotropy profile against the available data, on the large (left) and small (right) radial scale. Solid lines represent the model profiles, open circles the observational data points from \citet{vanLeeuwen2000} and black dots the data from \citet{Anderson2010}. Vertical bars indicate the measured errors and horizontal bars indicate the size of the bins.}
\label{fig:pm_omega}
\end{figure*}

For each selected model that meets the kinematic criteria we calculated the physical scales by means of the fits described in Sect. \ref{physical} and Sect. \ref{distance}. The best-fit model is taken to be the one that minimizes the total chi squared (defined as the sum of the calculated chi squared for the photometric, kinematic, and distance fits). As an example of this procedure, in Fig. \ref{fig:fit_omega} we show three different models for $\omega$ Cen, characterized by different values of the $V^{\mathrm{rot}}_{\mathrm{max}}/\sigma_0$ parameter, respectively 0.28, 0.34, and 0.36. The three models give comparable results for the kinematic profiles, very similar results for the photometric profile, and, most importantly, they all give similar trends in the predicted ellipticity profiles, as shown in Fig. \ref{fig:ell_omegacen}. Therefore we conclude that our selection procedure can be considered to be sufficiently stable and reliable.

Moreover, we recall that the kinematic data on the plane of the sky for 47~Tuc and M15 (see Sect. \ref{47} and \ref{M}, respectively) are not radially extended enough to allow us to determine the complete shape of the anisotropy profile. Therefore, in these cases the $R_\mathrm{a}$ scale, which marks the radial position of the transition from radial to tangential anisotropy, cannot be used as an additional criterion for the selection of the dimensionless parameters. However, the exploration of the parameter space just described already includes models with varying $R_\mathrm{a}$, because the shape of the anisotropy profile is directly intertwined with the observational parameters taken into consideration.

\section{$\omega$ Cen}
\label{omega}
The globular cluster $\omega$ Cen is the cluster for which the most complete photometric and kinematic data are available. In particular, the data set considered in this paper consists of 1\,868 line-of-sight velocities, 2\,740 ground-based proper motions, and 72\,970 HST proper motions (see Appendix \ref{App.A}). The kinematic profiles that we have constructed from these data extend out to $\approx0.5r_\mathrm{tr}$; therefore, a thorough comparison between models and observations can be carried out.
\begin{figure*}[t]
\centering
\includegraphics[width=0.80\textwidth]{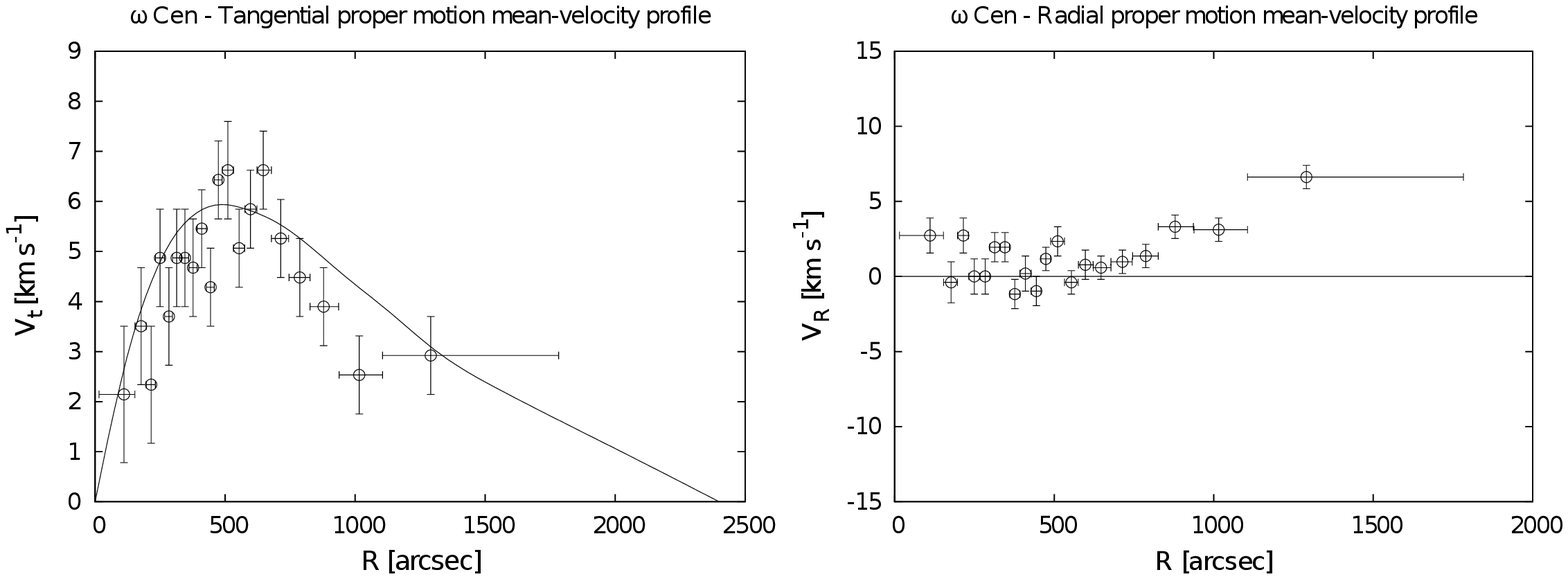}
\caption{Predicted proper motion mean-velocity profiles along the tangential and radial directions for $\omega$ Cen. Solid lines represent the model profiles and open circles the observational data points. Vertical bars indicate the measured errors and horizontal bars the size of the bins. Note that the data give a hint of a possible overall expansion, which is obviously not present in the model.}
\label{fig:pm_rot_omega}
\end{figure*}

\subsection{Photometry and kinematics}
\label{omega fit}

In general, the selected model is in satisfactory agreement with the surface brightness profile and the line-of-sight kinematic profiles, as shown in Fig.~\ref{fig:fit_omega}. For the photometric profile, the model reproduces well the central regions and the intermediate parts, but it underestimates the last two data points. For the line-of-sight kinematic profiles, the model is able to reproduce simultaneously the shape of the rotation profile and the shape of the velocity dispersion profile, with one important failure: the central values (inside $\approx 200$ arcsec) of the line-of-sight velocity dispersion are severely underestimated by our model. It is interesting to note that any quasi-Maxwellian dynamical model applied to $\omega$ Cen is unable to reproduce the cuspy behavior observed in the central regions (e.g., see the application of spherical King models and of spherical Wilson models presented by MLvdM05 in their Fig.~11; see also the fit by means of the rotating \citealt{Wilson1975} model performed by \citealp{Sollima2009}). In this respect, radially-biased anisotropic models appear to perform better (in particular, see the application of the $f^{(\nu)}$ models discussed by ZBV12). On the one hand, this feature has sometimes been considered as evidence for the presence of a central IMBH \citep[see][]{Noyola2008}. On the other hand, the same feature may indicate that $\omega$ Cen, because of its relatively high relaxation times (see Table~1), is only partially relaxed and characterized by a higher degree of radial anisotropy with respect to the case of more relaxed stellar systems, as suggested by Fig.~\ref{fig:pm_omega} (see also \citealp{vdMA2010}). A more detailed discussion of this issue is postponed to Sect. \ref{comparison}, where a comparison among models with different anisotropy profiles is presented. Curiously, even though (see Appendix \ref{App.B}) the line-of-sight data indicate high rotation in the very central regions ($R<0.5R_\mathrm{c}$), which is naturally interpreted as the signature of a complex rotating central structure, this does not appear to affect the quality of our results on the rotation profile; in fact, the selected model reproduces the central part of the line-of-sight rotation curve surprisingly well (see Fig.~2). 

In addition, the model identified by our procedure is able to reproduce all three components of the projected velocity dispersion tensor (both along the line-of-sight and on the plane of the sky; see Fig.~\ref{fig:pm_omega}). Interestingly, the shape of the observed anisotropy profile built from the proper motion dispersions is consistent with the general properties of the selected model, which is characterized by isotropy in the central region, weak radial anisotropy in the intermediate region, and tangential anisotropy in the outer parts. The transition between the region characterized by radial anisotropy to the region characterized by tangential anisotropy takes place at $R\approx1200$ arcsec. The data indeed show signs of radial anisotropy in the intermediate region (note that our model predicts a degree of radial anisotropy lower than the one observed) and of tangential anisotropy outside $R\approx 1000$ arcsec. The existence of tangential anisotropy found in the present study is consistent with the results of previous investigations, namely \citet{vandeVen2006} and \citet{vdMA2010} (see their Fig. 6). We wish to emphasize that such behavior of the anisotropy profile in the outer parts is a natural property of the family of models at the basis of the present work.

Finally, we can also compare the rotation on the plane of the sky predicted by the model with the observed mean-velocity profiles along the tangential and radial directions. Figure \ref{fig:pm_rot_omega} shows that the tangential proper motion mean-velocity profile $V_t(R)$ is well reproduced by the model, confirming the presence of differential rotation. In the radial direction the model predicts a flat profile with vanishing velocity; in the external regions ($R>1000$ arcsec), the observed proper motion mean-velocity in the radial direction reaches a value of $V_\mathrm{R}\approx5$ km s$^{-1}$, indicating the presence of a systematic expansion ascribed to systematic errors in the measurement procedures \citep{vanLeeuwen2000}. At this point, we should also recall that the procedure used to measure the proper motions removes any sign of solid-body rotation in the plane of the sky; therefore we apply to the data a correction to recover the solid-body mean velocity component, following \citet{vandeVen2006}, as discussed in Appendix \ref{App.A}. This fact introduces some uncertainties in the final profiles and might account for some of the discrepancies between the model and the observed proper motion mean-velocity profiles.

In conclusion, aside from the inner cusp problem, the generally good agreement between model and proper motion mean-velocity and velocity dispersion profiles is quite remarkable, because the model was selected only to match the velocity-to-dispersion ratio along the line-of-sight $V^{\mathrm{rot}}_{\mathrm{max}}/\sigma_0$, the location of the peak in the rotation profile along the line-of-sight $R^{\mathrm{rot}}_{\mathrm{max}}$, and the location of the transition from radial to tangential anisotropy in the plane of the sky.

\begin{figure}[b]
\centering
\includegraphics[width=0.45\textwidth]{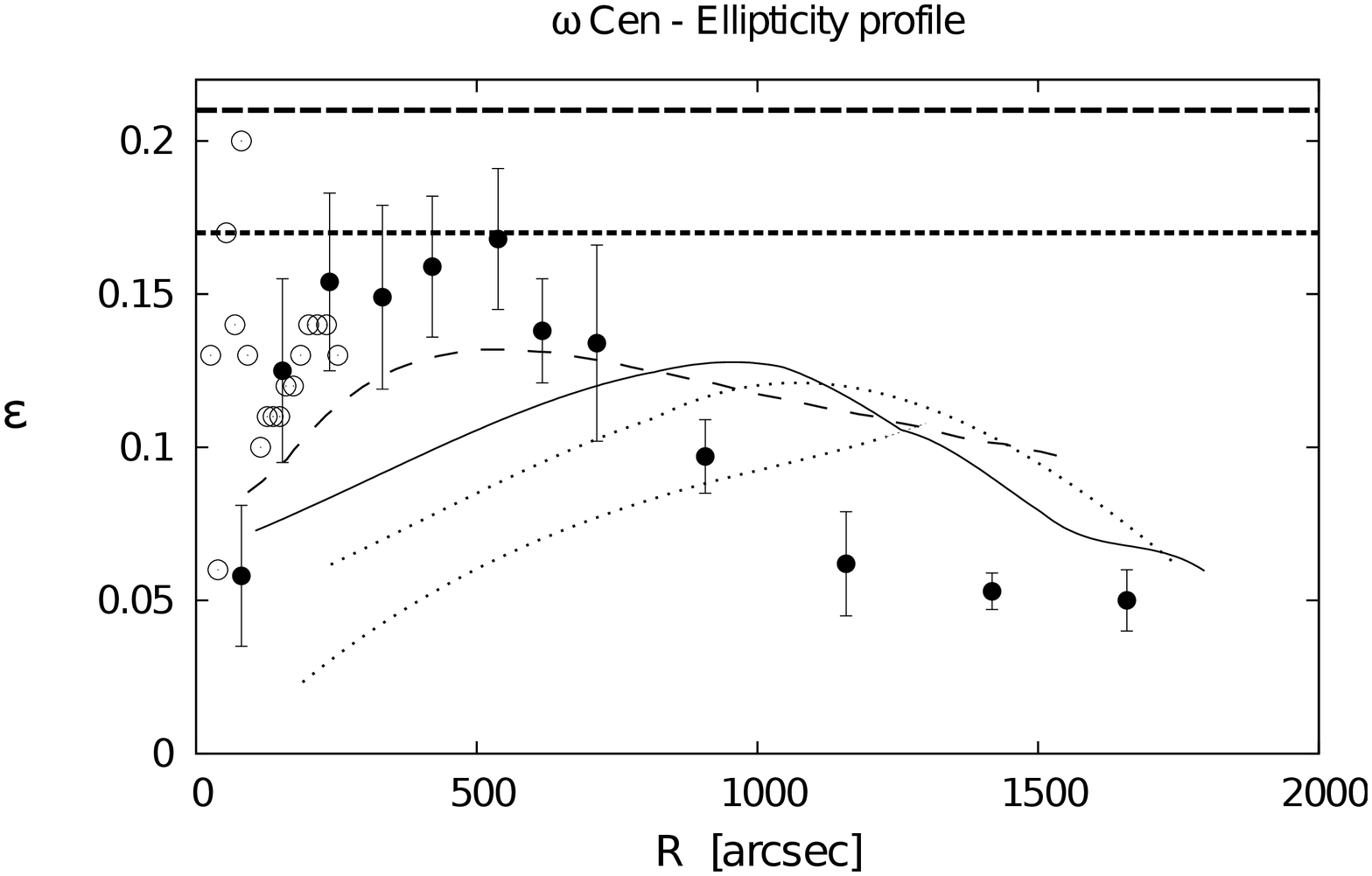}
\caption{Ellipticity profile for $\omega$ Cen. Open circles mark the observed ellipticities from \citet{Anderson2010}, black dots those from \citet{Geyer1983}. The solid line represents the predicted profile derived from the rotating axisymmetric model proposed in this paper, whereas the thin dotted curves correspond to the models used to test the sensitivity of the selection procedure (see Sect. \ref{exploration}). Dotted and dashed horizontal lines indicate the average values from WS87 and CC10, respectively.  Finally, the long-dashed line represents the ellipticity profile for the best-fit rotating Wilson (1975) model, from \citet{Sollima2009}; see discussion in Sect. \ref{sollima}.}
\label{fig:ell_omegacen}
\vspace{0.3cm}
\includegraphics[width=0.35\textwidth]{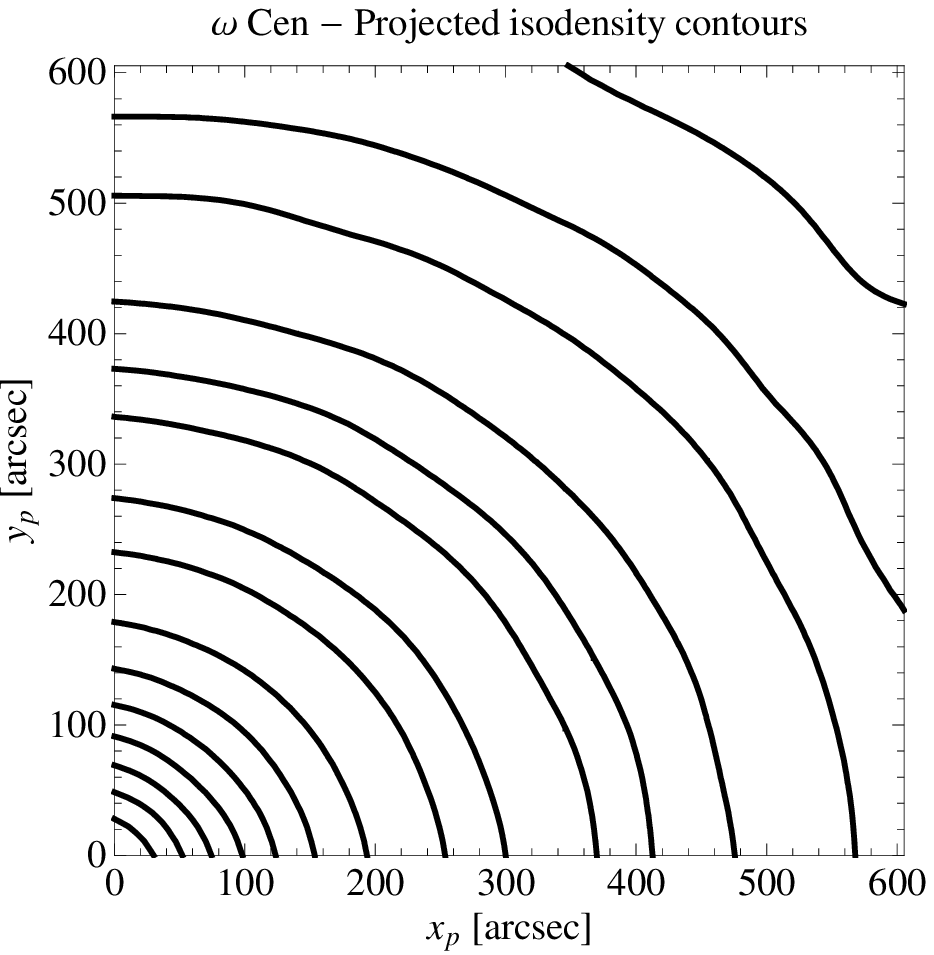}
\caption{Predicted projected isodensity contours for $\omega$ Cen. The contours are calculated in the first quadrant of the plane of the sky and correspond to selected values of the projected number density (normalized to the central value) in the range $[0.9, 10^{-2}]$. The area represented in the figure covers a square of side length approximately equal to 2$R_h$.}
\label{fig:iso_omegacen}
\end{figure}

\subsection{Dynamical distance}
\label{omega distance}
The rescaling of the model profiles to match the observed proper motion dispersion profiles allows us to derive an estimate for the distance of the cluster (see Sect.~\ref{distance}). The dynamical distance obtained for $\omega$ Cen is $d = 4.11\pm 0.07$ kpc, with an associated reduced chi-squared $\widetilde{\chi}^2_{\mathrm{d}}=7.19$. This value is significantly smaller than the distance estimated with photometric methods (e.g., $d=5.2\pm0.7$ kpc from \citealp{HarrisCat2010}) and also smaller than other estimates obtained by means of the application of different dynamical models (e.g., $d=4.70\pm0.06$ kpc from \citealp{vdMA2010}; $d=4.8\pm0.3$ kpc from \citealp{vandeVen2006}).

As also noted by \citet{vandeVen2006}, a low value of the distance is expected when either the line-of-sight velocity dispersion is underestimated or the proper motion dispersion is overestimated. In our case, it is clear from the previous section and from Fig.~\ref{fig:fit_omega}, that our dynamical model underestimates the central value of the line-of-sight velocity dispersion. Therefore, our distance estimate is affected by a systematic bias (reflected also by the high value of the reduced chi-squared). The dynamical distances obtained by \citet{vdMA2010} and \citet{vandeVen2006} are based on a Jeans model and on an orbit-based model, respectively; previous studies based on the application of quasi-Maxwellian dynamical models, such as spherical King or spherical Wilson models, do not report distance estimates for this object.

\subsection{Deviations from spherical symmetry}
\label{ellipticity_omega}

The selected axisymmetric model is associated with a well defined ellipticity profile, which is the morphological counterpart to the presence of rotation. The comparison with the corresponding observed profile is illustrated in Fig.~\ref{fig:ell_omegacen}; the open circles represent the profile from \citet{Anderson2010}, the black dots represent the profile from \citet{Geyer1983}, and the solid line the profile derived from our model. The two observed profiles are consistent in the sampled radial range, except for the innermost region ($R<100$ arcsec) where a large scatter dominates the data of \citet{Anderson2010}. For completeness, in Fig.~\ref{fig:iso_omegacen} we present the projected isodensity contours predicted by our model, which show the two-dimensional deviations from sphericity. The contour shapes are of interest for future comparisons with observations based on more detailed morphological studies and may provide an important clue to distinguish between different dynamical models (e.g., see VB12).

\begin{figure*}[p]
\centering
\includegraphics[width=0.80\textwidth]{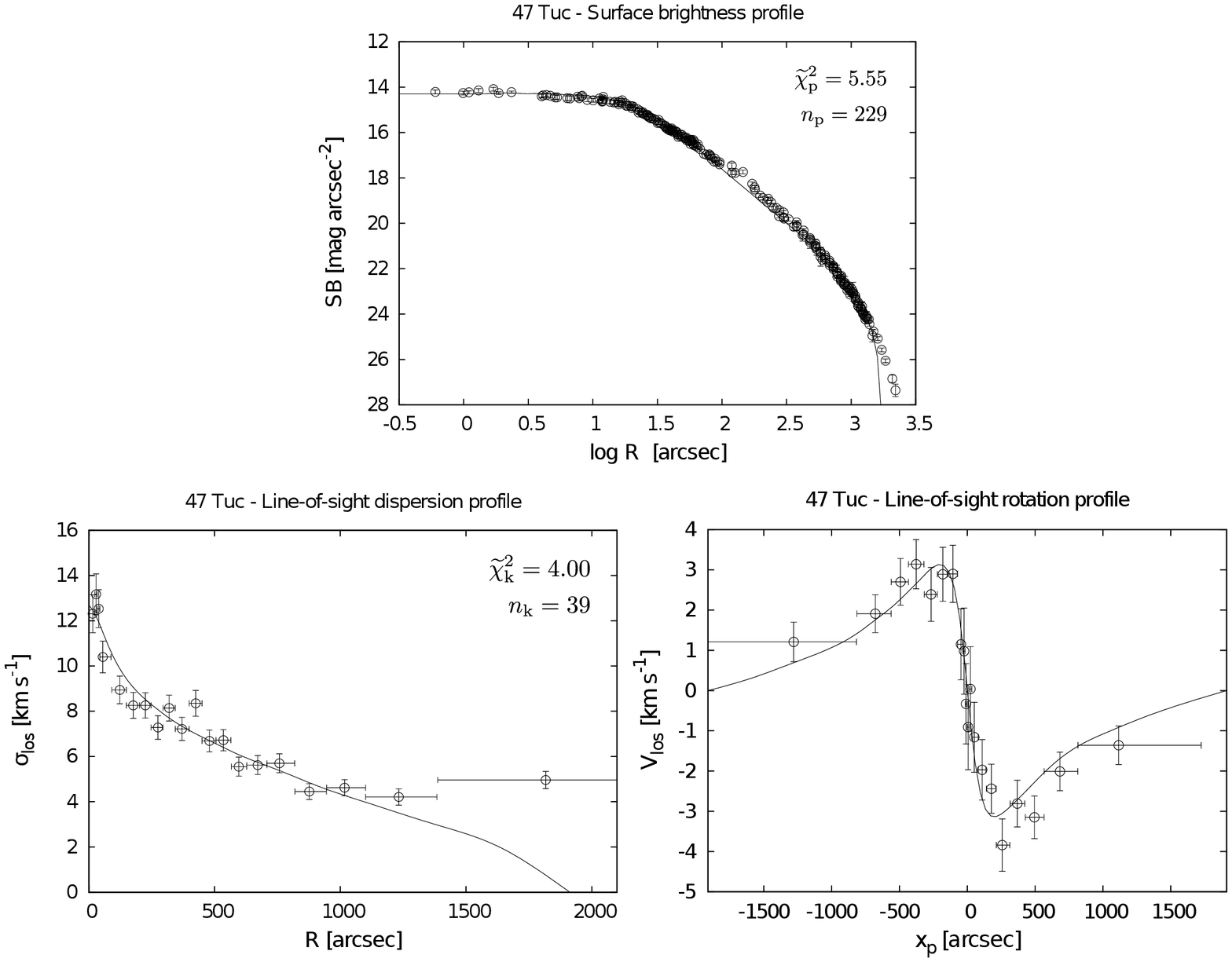}
\caption{Surface brightness profile, line-of-sight velocity dispersion profile, and line-of-sight rotation profile (measured along the projected major axis) for 47 Tuc. The associated photometric and kinematic reduced chi-squared and the number of degrees of freedom are shown. For description of symbols and curves see Fig. \ref{fig:fit_omega}.}
\label{fig:fit_47tuc}
\vspace{0.5cm}
\includegraphics[width=0.80\textwidth]{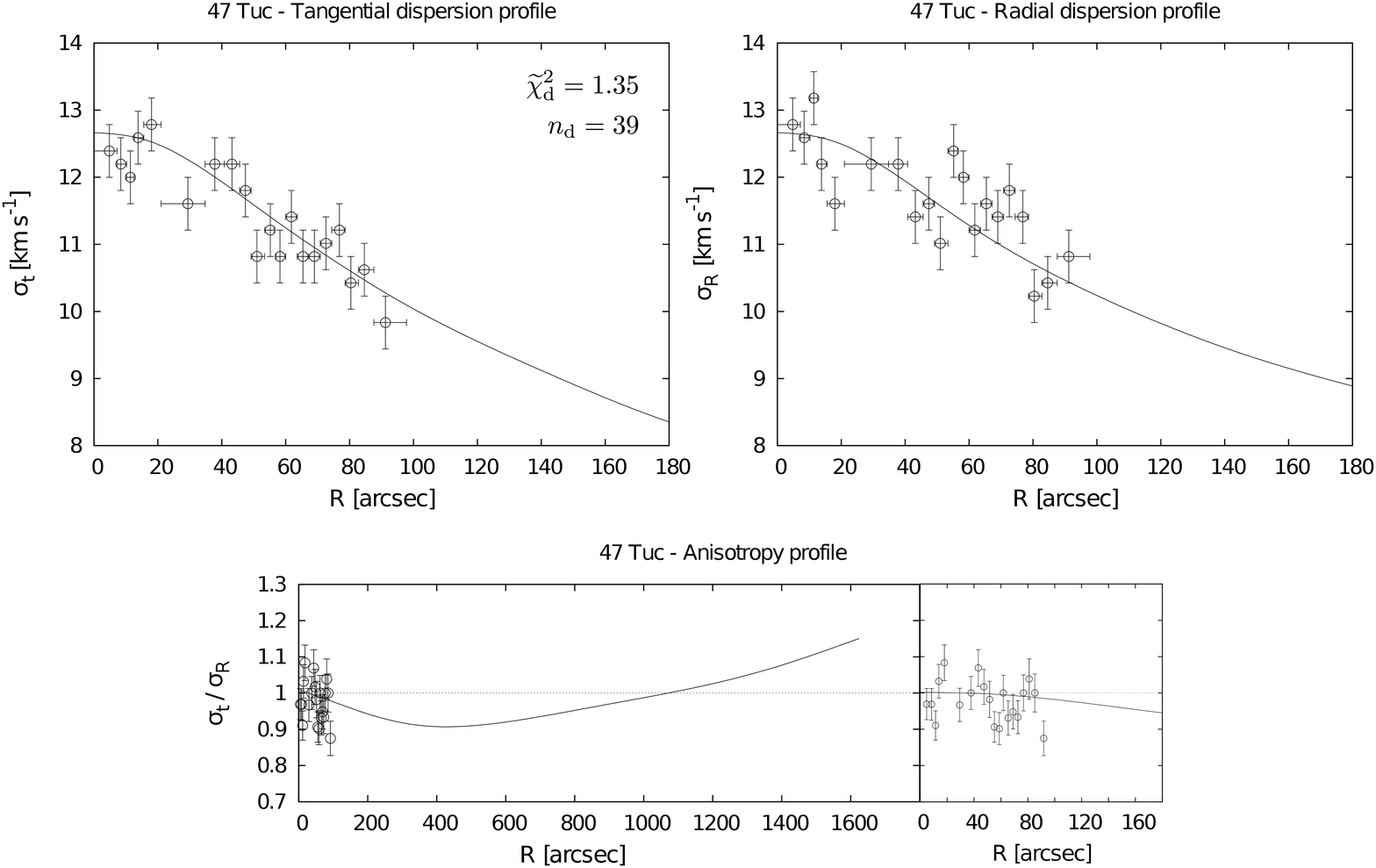}
\caption{Tangential dispersion, radial dispersion, and anisotropy profile for 47 Tuc. The associated reduced chi-squared and the number of degrees of freedom are shown. For description of symbols and curves see Fig. \ref{fig:pm_omega}.
}
\label{fig:pm_47tuc}
\end{figure*}

The model ellipticity profile is characterized by a general trend similar to that of the \citet{Geyer1983} measurements, but it predicts the peak of maximum flattening too far out, at about $R\approx1000$ arcsec. If we calculate the average ellipticity in the radial range covered by the data, we find an average flattening associated with the selected model ($\varepsilon=0.10$) in agreement with the observed one ($\varepsilon=0.12\pm0.02$). In other words, we are led to conclude that the observed deviations from sphericity are likely to be originated by the presence of internal rotation. In Sect. \ref{comparison} we will argue that the discrepancy between the predicted and observed ellipticity profiles are likely to be related to the complex nature of $\omega$ Cen, in particular to its conditions of partial relaxation and the interplay between rotation and anisotropy in velocity space. In this respect, we expect that our models of quasi-relaxed stellar systems perform better for globular clusters characterized by shorter relaxation times (such as 47 Tuc and M15).

\section{47 Tuc}
\label{47}

For the globular cluster 47 Tuc the data set consists of 2\,476 line-of-sight-velocities and 12\,974 HST proper motions (see Appendix \ref{App.A}). The line-of-sight kinematical data cover the full radial extent of the cluster, out to approximately the truncation radius. In turn, the proper motion data are limited to a disk of radius $4R_\mathrm{c}$.

\subsection{Photometry and kinematics}
As illustrated in Fig.~\ref{fig:fit_47tuc}, the surface brightness profile and the line-of-sight rotation and velocity dispersion profiles are well reproduced by the selected model. In particular, the rotation profile is well matched throughout the extension of the cluster, showing clearly the position of the maximum rotation velocity, the characteristic rigid rotation behavior in the central region, and the relatively sharp decrease in the outer parts. The observed line-of-sight velocity dispersion profile is characterized by one data-point at $R\ga1800$ arcsec deviating from the model profile. A corresponding discrepancy is found also for the surface brightness profile, at approximately the same radial position (the last four photometric data-points). These two features may be interpreted in terms of the population of ``potential escapers''  resulting from the tidal interaction between the cluster and the host Galaxy \citep[see][]{Kupper2010,Lane2012}.

As to the proper motion data, the relevant profiles, although limited to the central region, show a satisfactory agreement with the model predictions (see Fig.~\ref{fig:pm_47tuc}). In the intermediate regions ($50\la R \la1000$~arcsec) the model predicts weak radial anisotropy and tangential anisotropy in the outer parts. It would be interesting to acquire more spatially extended proper motion measurements to confirm this prediction [in line with the results obtained for the anisotropy profile of $\omega$ Cen (see Fig.~\ref{fig:pm_omega})].

Rotation in the plane of the sky is not available from the proper motion data set of \citet{McLaughlin2006}. However, proper motion rotation has been measured by \citet{Anderson2003}, by using the HST and by considering background stars of the Small Magellanic Cloud as an absolute reference frame. The observed rotation corresponds to a velocity of $4.97\pm1.17$ km s$^{-1}$ (based on the assumed distance of 4.5 kpc) at a projected radius of 5.7 arcmin (corresponding approximately to the position of the rotation peak). Within the uncertainties, this is consistent with our model, which predicts a value of 4.13 km s$^{-1}$ at 5.7 arcmin.

\subsection{Dynamical distance}
The comparison of the observed proper motion dispersion profiles with the model predictions allows us to derive an estimate of the distance to the cluster (see Sect. \ref{distance}). For 47 Tuc the best-fit distance is $d=4.15\pm0.07$ kpc, with associated reduced chi-squared $\widetilde{\chi}^2_{\mathrm{d}}=1.35$ inside the corresponding 90\% CI. This value is consistent with the dynamical distance reported by \cite{McLaughlin2006} $d=4.02\pm0.35$ kpc, measured from the same proper motion data set used in the present work, under the simple assumptions of spherical symmetry, isotropy, and absence of internal rotation. Our value is lower than the standard value of $d=4.5\pm0.2$ kpc reported in the Harris catalog \citep{HarrisCat2010} and lower than other distance estimates obtained by means of photometric methods, such as main sequence fitting, RR Lyrae, and white-dwarf cooling sequence fitting (for a recent summary of results, see Table 1 of \citealp{Woodley2012} or \citealp{Bono2008}).

\subsection{Deviations from spherical symmetry}
\label{47ell}

\begin{figure}[t]
\centering
\includegraphics[width=0.45\textwidth]{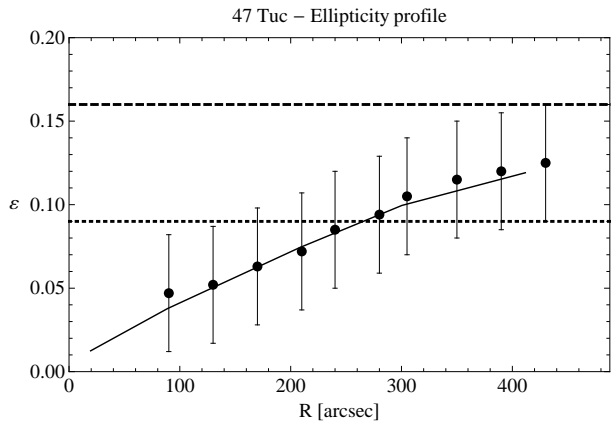}
\caption{Ellipticity profile for 47 Tuc. The black dots mark the observed ellipticities presented by WS87, the solid line represents the profile derived from our axisymmetric rotating model. Dotted and dashed horizontal lines indicate the average values reported by WS87 and CC10, respectively.}
\label{fig:ell_47tuc}
\end{figure}

Figure \ref{fig:ell_47tuc} shows the ellipticity profile predicted by our model plotted together with the ellipticity data available for 47 Tuc. In this cluster, the deviations from spherical symmetry are naturally explained by the selected model with a surprising degree of accuracy. In fact, the ellipticity profile derived by our model reproduces the radial variation of the observed ellipticity over the entire spatial range covered by the data (the flattening of 47 Tuc increases from a value of $\varepsilon\approx0$ to a maximum value of $\varepsilon\approx0.12$ at $R\approx450$ arcsec). We recall that the ellipticity profile associated with the selected self-consistent model is a structural property completely determined by the dimensionless parameters and physical scales identified during the model selection procedure. In this case we can thus state with confidence that internal rotation is the physical ingredient responsible for the observed global deviations from spherical symmetry. In this respect, we emphasize that the relation between the shapes of the rotation profile and the ellipticity profile is highly nontrivial; in particular, the peak of the rotation profile does not correspond to a peak in the ellipticity profile (at variance with what is often believed, e.g. \citealp{Meylan1986}).

\section{M15}
\label{M}

\begin{figure*}[!ph]
\centering
\includegraphics[width=0.80\textwidth]{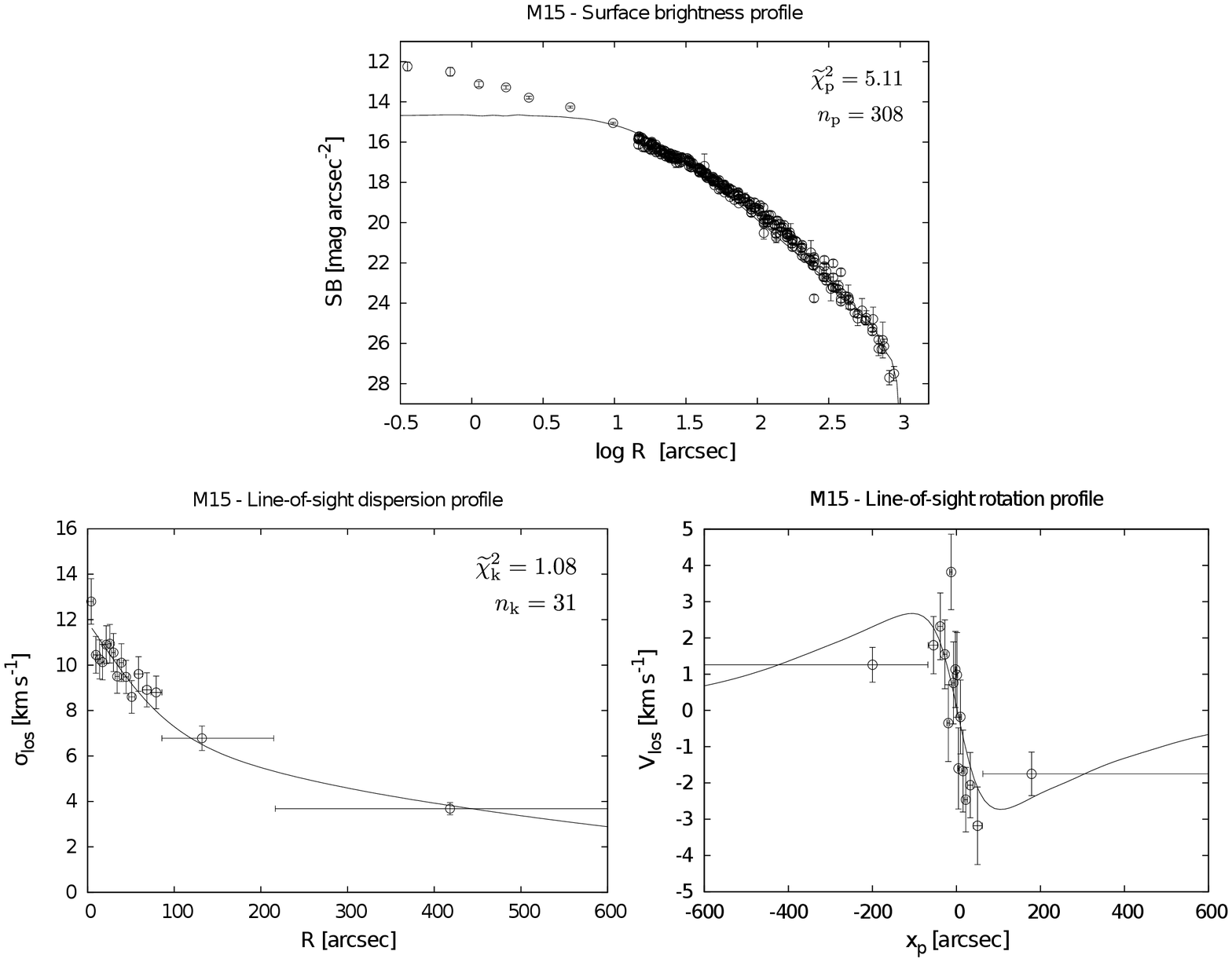}
\caption{Surface brightness profile, line-of-sight velocity dispersion profile, and line-of-sight rotation profile (measured along the projected major axis) for M15. The associated photometric and kinematic reduced chi-squared and the number of degrees of freedom are shown. For description of symbols and curves see Fig. \ref{fig:fit_omega}.
}
\label{fig:fit_M15}
\vspace{0.5cm}
\includegraphics[width=0.80\textwidth]{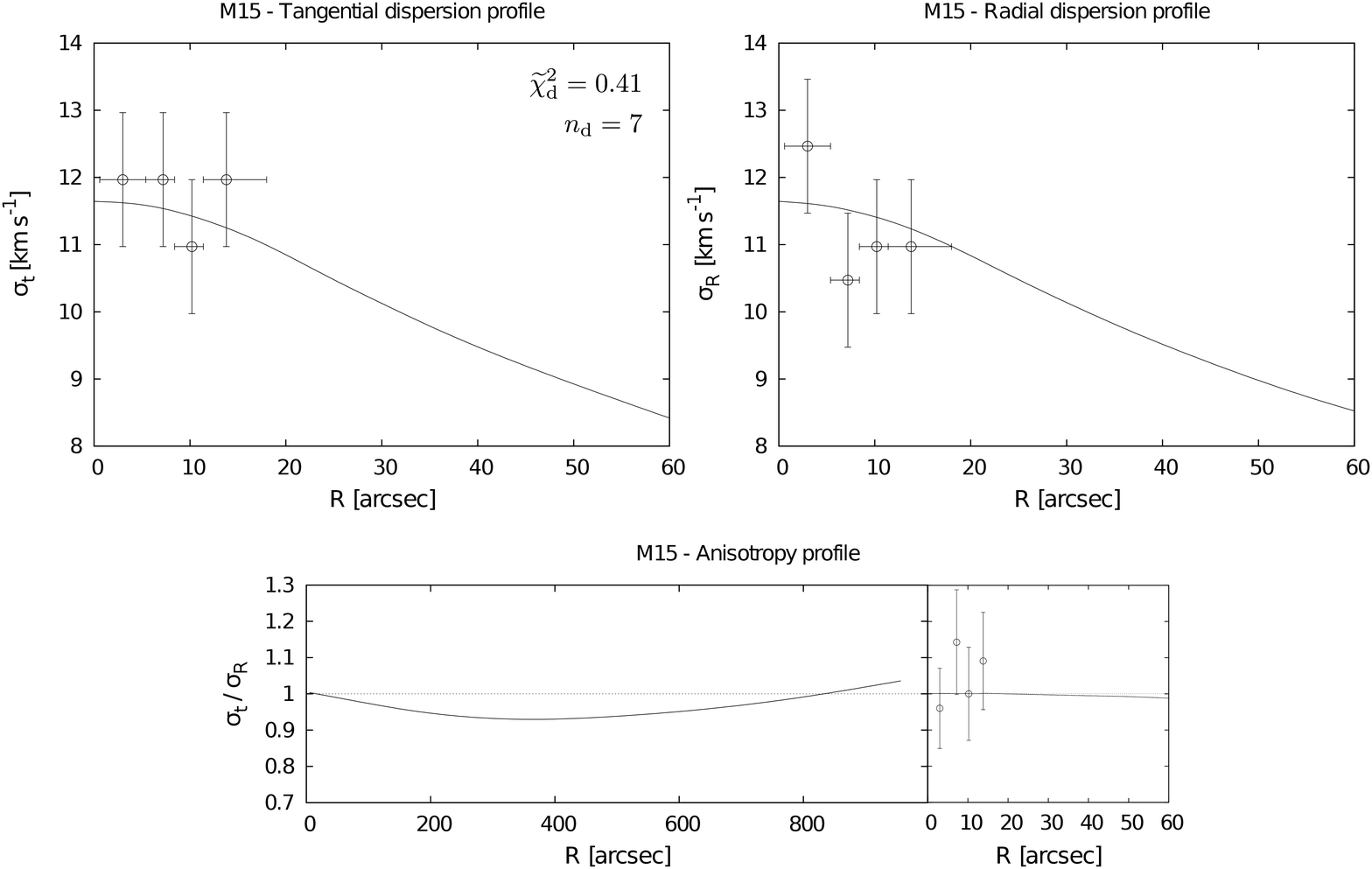}
\caption{
Tangential dispersion, radial dispersion, and anisotropy profile for M15. The associated reduced chi-squared and the number of degrees of freedom are shown. For description of symbols and curves see Fig. \ref{fig:pm_omega}.}

\label{fig:pm_M15}
\end{figure*}

The studies of the globular cluster M15 are largely focused on its central region. In fact, the cluster is believed to be in a post-core-collapse phase and mass segregation is thought to play a role in its dynamics. In particular, the sharp gradient of the central luminosity is thought to be the result of the dynamical evolution of the cluster (e.g., see \citealp{Baumgardt2003} and \citealp{Murphy2011}) or of the presence of a central intermediate mass black hole (e.g., see \citealp{Gerssen2002}). The available kinematic data are limited to the central regions and consist of 1\,777 line-of-sight velocities and 703 HST proper motions (see Appendix \ref{App.A}).

\subsection{Photometry and kinematics}
Remarkably, except for the most central region, the selected model offers a good description of both the line-of-sight kinematic profiles and the surface brightness profile (see Fig.~\ref{fig:fit_M15}). The line-of-sight velocity dispersion profile is reproduced by the model out to the last available bin, located at approximately $0.5 r_\mathrm{tr}$.

As to the line-of-sight rotation profile, a large scatter is present in the central regions, due to the high measurement errors, which have an average of 3.79 km s$^{-1}$ (significantly higher than the average errors of $\omega$ Cen and 47 Tuc: 1.98 km s$^{-1}$ and 2.29 km s$^{-1}$, respectively). Unfortunately, the kinematic data set does not sample the region where the peak of the rotation curve is expected. More accurate and better distributed line-of-sight velocity measurements would be required to build a more reliable and complete rotation profile. However, it is interesting to note that the rotation profile in the central regions, characterized by a solid-body behavior, is well accounted for by the model, although high rotation is detected in the center and interpreted as a signature of the presence of a decoupled rotating core (see Appendix \ref{App.B}).

For the proper motions, given the small number of data and the low accuracy of the measurements, we decided to divide the sample in only 4 bins to avoid excessive statistical noise; the relevant profiles are illustrated in Fig.~\ref{fig:pm_M15}. Such profiles can be used to constrain the kinematic behavior of the cluster only in relation to the very central regions. In turn, the selected model leads to specific predictions on the anisotropy profile in the intermediate and outer parts of the object, which are expected to first show weak radial anisotropy and then tangential anisotropy. Unfortunately, for this object no information about the rotation on the plane of the sky is yet available.

\begin{figure}[t]
\centering
\includegraphics[width=0.45\textwidth]{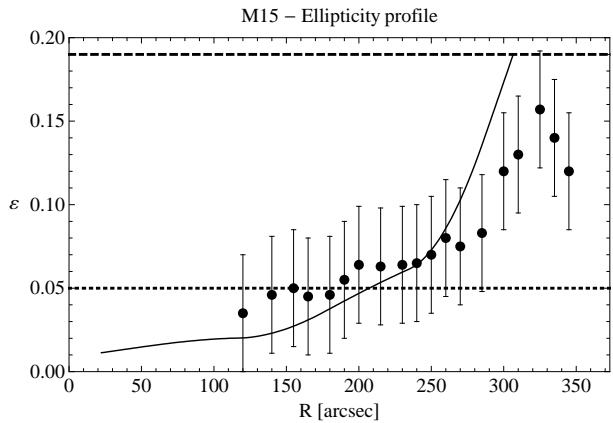}
\caption{Ellipticity profile for M15.  For description of symbols and curves see Fig. \ref{fig:ell_47tuc}.}
\label{fig:ell_m15}
\end{figure}

\subsection{Dynamical distance}
The dynamical distance obtained from the procedure described in Sect. \ref{distance} yields a distance of $d=10.52\pm0.38$~kpc, with a reduced chi-squared $\widetilde{\chi}^2_{\mathrm{d}}=0.41$, inside the corresponding 90\% CI. This is consistent with the kinematic distance obtained by \citet{McNamara2004} of $d=9.98\pm0.47$ kpc and the value obtained by \citet{vandenBosch2006} of $d=10.3\pm0.4$ kpc; these two estimates are based on the same proper motion data set considered in the present work. In particular, the value obtained by \citet{McNamara2004}, which is based on the simplifying assumptions of spherical symmetry, isotropy in velocity space, and no rotation, is lower than the value obtained in the present paper and the one obtained by \citet{vandenBosch2006} (in which anisotropy, rotation, and flattening are taken into account). Moreover, our distance is in agreement with other distance estimates based on photometric methods, such as the one reported in the Harris catalog \citep{HarrisCat2010} $d=10.4\pm0.8$ kpc. In this case, the conclusion drawn by \citet{Bono2008}, according to which distances obtained from kinematic data are systematically lower than distances obtained from other methods, does not hold.

\subsection{Deviations from spherical symmetry}
\label{Mell}

The comparison between the observed and the predicted ellipticity profiles is illustrated in Fig.~\ref{fig:ell_m15}. Our model predicts a value of ellipticity close to zero in the very central regions and an increase of the flattening thereafter, consistent with the observations. In particular, we note that the model profile seems to overlap smoothly with the observed profile in the region sampled by the data. Moreover, the observed average flattening is consistent with the value predicted by our model. We thus conclude that our dynamical model, and consequently the presence of internal rotation, can naturally explain the observed deviations from sphericity of M15.

\section{Discussion}
\label{comparison}

\subsection{Partially relaxed vs. well-relaxed clusters}
\label{sollima}
The three globular clusters under consideration are known to be in different evolutionary states. In fact, the core relaxation time of $\omega$ Cen is significantly higher than the relaxation times of 47 Tuc and M15 (see Table~\ref{tab:1}). This suggests that $\omega$~Cen should be in a partially relaxed state, whereas 47 Tuc and M15 can be considered well-relaxed clusters. 

\begin{figure}[t]
\centering
\includegraphics[width=0.45\textwidth]{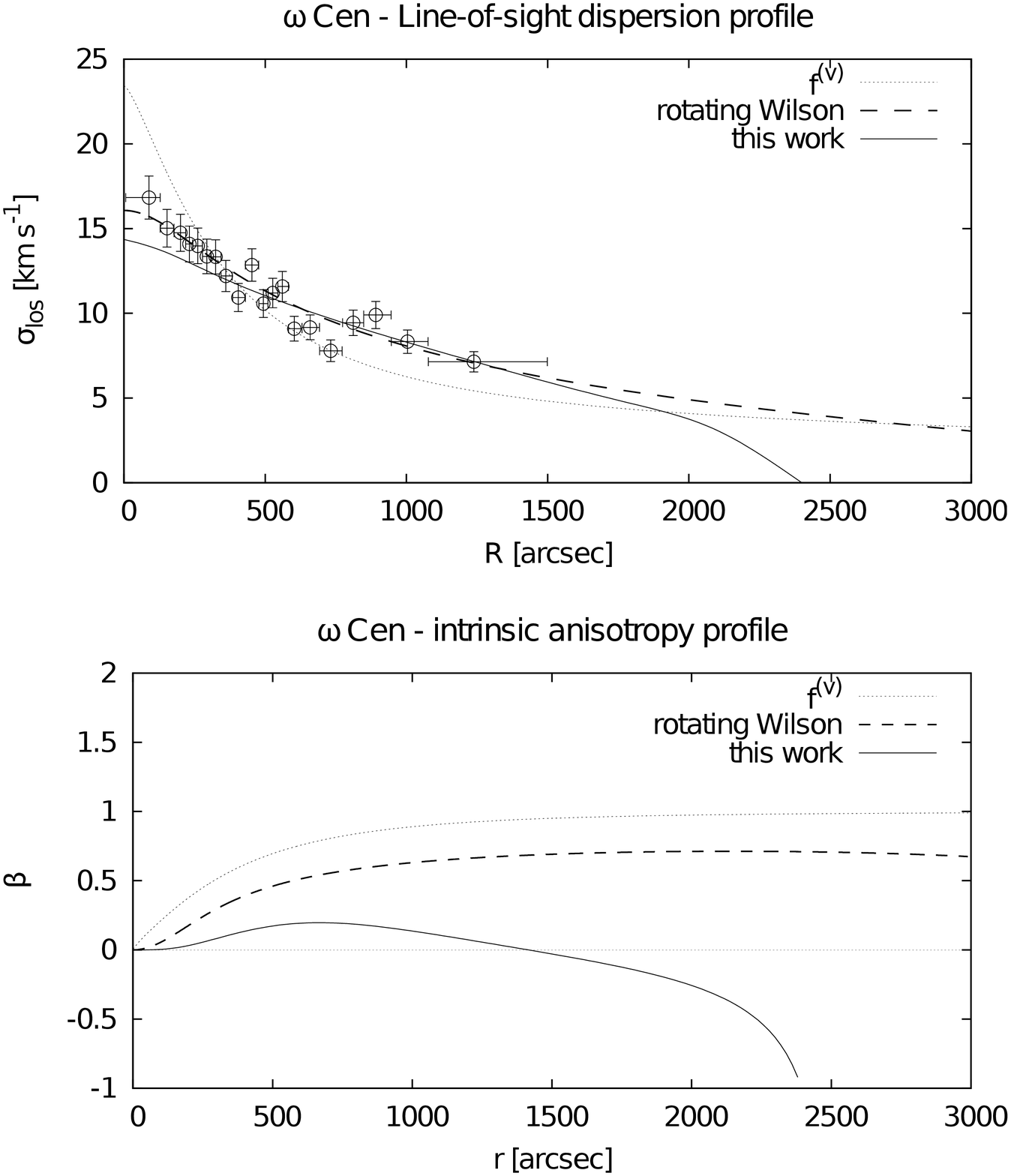}
\caption{Comparison between our rotating quasi-relaxed model (solid lines), spherical radially-biased anisotropic $f^{(\nu)}$ model (from ZBV12, thin dotted lines), and rotating \citet{Wilson1975} model \citep[from][dashed lines]{Sollima2009} for $\omega$ Cen. The top panel represents the projected line-of-sight velocity dispersion profile and the bottom panel the intrinsic anisotropy profile defined as $\beta=1-(\sigma^2_{\theta}+\sigma^2_{\phi})/2\sigma^2_r$, evaluated along the equatorial plane. A higher degree of radially-biased anisotropy in the central-intermediate region contributes to steepen the central dispersion profile.}
\label{fig:ani_comp}
\end{figure}

In the case of $\omega$ Cen, we argue indeed that the main discrepancies noted between our model and observations are associated with the condition of partial relaxation of the cluster. Our model is unable to describe the cuspy behavior of the velocity dispersion profile in the central regions (inside $\approx$ 300 arcsec). In Fig. \ref{fig:ani_comp} we compare our quasi-relaxed model with the best-fit (spherical, nonrotating, nontruncated) $f^{(\nu)}$ model from ZBV12 and the best-fit axisymmetric, rotating \citet{Wilson1975} model from \citet{Sollima2009}. The top panel shows that, in the central region ($R\lesssim500$~arcsec $\approx2R_\mathrm{h}$), the gradient of the line-of-sight dispersion profile depends strongly on the assumed model: the steepest gradient is associated with the spherical $f^{(\nu)}$ model, which is the model characterized by the strongest radial anisotropy. Note that the spherical $f^{(\nu)}$ model and the rotating \citet{Wilson1975} model both miss the feature of tangential anisotropy in the outer regions altogether. This is further illustrated by the bottom panel, which shows the intrinsic anisotropy parameter $\beta=1-(\sigma^2_{\theta}+\sigma^2_{\phi})/2\sigma^2_r$ profile evaluated along the equatorial plane. Indeed, the rotating models constructed in VB12 and applied in this paper are characterized by isotropy in the central regions and only weak radial anisotropy in the intermediate radial range, because they assume that the stellar system is quasi-relaxed.

In Fig. \ref{fig:ell_omegacen} the ellipticity profile predicted by our model is compared to the profile presented by \citet{Sollima2009} based on a rotating \citet{Wilson1975} model: the latter model
generates deviations from sphericity in better agreement with the observations in the inner regions, but not in the outer parts (beyond $\approx$ 1000 arcsec), where the model is radially anisotropic, whereas $\omega$ Cen is tangentially anisotropic. 

We conclude that the structure of $\omega$~Cen is determined by the complex interplay between rotation and anisotropy; significant pressure anisotropy can be naturally present even in its inner regions because this cluster is characterized by long relaxation times.

When applied to the two more relaxed clusters, 47~Tuc and M15, our models perform very well; the systems are quasi-isotropic in their inner regions and internal rotation is able to explain the observed morphology. The most significant discrepancy left is probably that of the core structure of M15 (inside $\approx$ 10 arcsec), characterized by a cusp in the surface brightness that is likely to be related to the phenomenon of core collapse \citep{Murphy2011}, which goes beyond the objectives of our equilibrium models. For this cluster, the intermediate and outer regions (from 10 arcsec out to 1000 arcsec) are well fitted by our rotating model (Fig.~\ref{fig:fit_M15}), at variance with the spherical King model, which severely underestimates the surface brightness (beyond $\approx$ 300 arcsec; see Fig.~1 in ZBV12).

\subsection{Comparison with previous studies}

To our knowledge, an application of nonspherical models to the full set of data available for these clusters, including proper motions, has been made only by \citet{vandeVen2006} for $\omega$ Cen and by \cite{vandenBosch2006} for M15, based on a Schwarzschild-type modeling procedure. Remarkably, the best-fit model for M15 is characterized by a total mass and a mass-to-light ratio fully consistent with our results, that is, $4.4\times10^5 M_\odot$ and 1.6 $M_\odot/L_\odot$, respectively. In the case of $\omega$ Cen, we derive a lower value for the total mass and a higher value for the mass-to-light ratio. Here the discrepancy reflects our estimate of the distance to the object, smaller than distances reported in the literature\footnote{For $\omega$ Cen, the recent investigation by \citet{Dsouza2012} assumes a distance of 5.5 kpc, much higher than the distance (4.11 kpc) that we determined in the present paper. Based on a discrete kinematic approach, including flattening and rotation, the authors report a value of the total mass of $(4.05\pm0.10)\times10^6M_\odot$. By assuming an apparent visual magnitude of $m_\mathrm{V,tot}=3.68$ mag \citep{HarrisCat2010}, and by rescaling this value to the distance of 5.5 kpc (to obtain the absolute total luminosity), the corresponding mass-to-light ratio is $M/L_\mathrm{V}=4.56\,M_\odot/L_\odot$, significantly larger than usually obtained for this cluster.} (by adopting a distance of $d=4.8$ kpc, the resulting total mass associated with our rotating model would be $M=2.28\times10^6M_\odot$, whereas for $d=5.2$ kpc, the total mass would be $M=2.47\times10^6M_\odot$).   

\begin{table}[t]
\caption[]{Comparison of the structural parameters from the best-fit models of the present paper with those obtained from spherical models in previous studies.}
\label{tab:7}
\centering
\begin{tabular}{cccccc}
\hline\hline
GC & Ref. & C & $R_\mathrm{c}$ & $M$ &   $M/L_\mathrm{V}$   \\
 \cline{5-6} 
       &	&	&			     &	   \multicolumn{2}{c}{\scriptsize{rescaled to a}} 	\\
       &	&	&			     &	   \multicolumn{2}{c}{\scriptsize{common distance}} 	\\

\hline
$\omega$ Cen  & (0)  & $1.27\pm0.01$ &  $127.8\pm1.1$& $24.71\pm0.20$& $2.26\pm0.11$\\
     & (1)  & $1.32\pm0.01$ & $127.7\pm2.4$ & $26.45\pm3.32$ & $1.93\pm0.24$   \\
     & (2)  & $1.43\pm0.02$  & $164.6\pm4.5$ & $24.66\pm2.26$&   $2.24^{+1.04}_{-0.82}$ \\
     & (3)  & $1.31\pm0.04$  & $142.2\pm8.3$ & $\ldots$ & $\ldots$    \\
47 Tuc  & (0)  &$1.87\pm0.01$  & $24.6\pm0.1$ & $6.76\pm0.04$ & $1.56\pm0.12$ \\
     & (1)  & $2.01\pm0.00$ & $22.6\pm0.2$  & $7.18\pm0.41$ &   $1.34\pm0.08$ \\
     & (2)  & $ 2.57\pm0.06$  & $32.1\pm2.6$ & $10.71\pm0.98$ &  $1.17^{+0.52}_{-0.43}$\\
     & (3)  &  $2.07\pm0.03$  & $21.6\pm1.3$ & $\ldots$ & $\ldots$    \\
M15 & (0)  & $1.94\pm0.02$ &  $12.9\pm0.2$ & $4.49\pm0.07$ & $1.47\pm0.05$  \\
     & (1)  &$1.86\pm0.01$  & $7.5\pm0.1$ & $3.98\pm0.35$ &  $1.12\pm0.10$  \\
     & (2)  &  $\ldots$  &  $\ldots$ & $\ldots$ &  $\ldots$  \\
     & (3)  &  $2.29\pm0.18$ & $8.4\pm1.0$ & $\ldots$ & $\ldots$    \\
\hline
\end{tabular}
\tablecomments{For each cluster we provide the concentration parameter $C=~\log(r_\mathrm{tr}/R_\mathrm{c})$, the projected core radius $R_\mathrm{c}$ in arcsec, the total mass of the cluster $M$ in units of $10^5M_\odot$, and the V-band mass-to-light ratio $M/L_V$ in solar units. The values of $M$ and $M/L_V$ have been rescaled to a common distance for each cluster to allow for a comparison of the different models considered (5.2 kpc, 4.5 kpc, and 10.4 kpc, for $\omega$ Cen, 47 Tuc, and M15, respectively).}
\tablerefs{(0) This paper; (1) spherical King models from ZBV12; (2) spherical nonrotating Wilson models from MLvdM05; (3) \citealp{HarrisCat2010}.}
\end{table}


In addition, only very few studies have been made of nonspherical rotating models constructed under given physical assumptions. To our knowledge, only three families of models based on a distribution function allowing for internal rotation have been explored in significant detail: those by \citet{Prendergast1970}, \citet{Wilson1975}, and \citet{LuptonGunn1987}. The first two were originally designed to describe elliptical galaxies and not globular clusters. In fact, the closest and modern paper that we are aware of, for which some comparison with the present article could be made, is that by Sollima et al. (2009), although the application presented there is limited to the line-of-sight kinematics (and thus without consideration of the star proper-motion data). The comparison was provided in the previous subsection.

Therefore, we are left with the task of comparing the results of the dynamical analysis performed in the present paper with the results obtained from previous studies based on spherical nonrotating models. This comparison is also interesting, because it shows to what extent the determination of the structural parameters is sensitive to the model adopted, or, in other words, to what extent some idealized, relatively simple,  commonly used models are likely to introduce systematic errors in probing the structure of globular clusters. Table \ref{tab:7} summarizes and compares the following derived structural properties: concentration parameter $C$, core radius $R_\mathrm{c}$, total mass $M$, and global V-band mass-to-light ratio $M/L_\mathrm{V}$. In general, the values of the derived structural parameters are consistent with the values derived from the other studies. Spherical nonrotating Wilson models tend to lead to larger truncation radii, as expected.

We note that our rotating models give a good description of the global kinematics and morphology of the three analyzed globular clusters. As a result, the effects of mass segregation are expected to be modest; in addition, we do not have to invoke the presence of dark matter and we do not find any reason to abandon Newtonian dynamics and to move to the MOND framework.
 
\subsection{$V/\sigma$ vs. $\varepsilon$}
\label{vsigma}

\begin{figure}[t]
\centering
\includegraphics[width=0.45\textwidth]{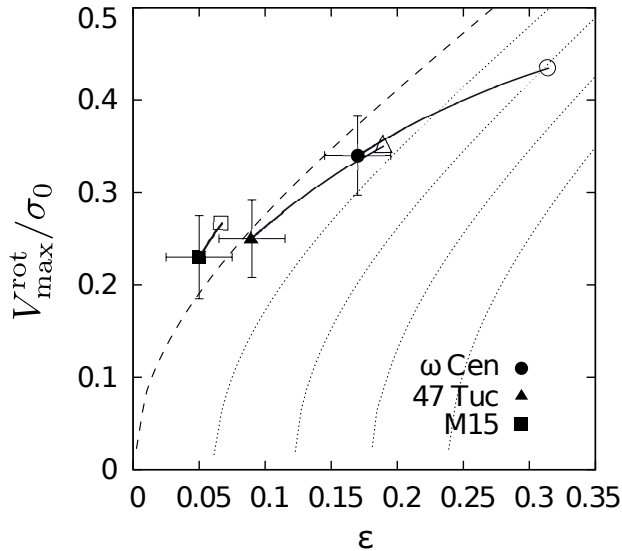}
\caption{$V/\sigma$ vs. ellipticity $\varepsilon$ for $\omega$ Cen, 47 Tuc, and M15. Filled symbols denote the pairs ($V/\sigma$,\,$\varepsilon$), in which the ellipticity values are determined by WS87. Empty symbols, connected by a segment to the associated filled symbols, indicate the pairs ($V/\sigma$,\,$\varepsilon$) corrected for inclination. The dashed line indicates the relation for isotropic oblate rotators viewed ``edge-on'', whereas the thin dotted lines indicate oblate rotators viewed ``edge-on'' with different global anisotropy parameter $\delta$ (from left to right $\delta=$ 0.05, 0.1, 0.15, 0.20). See text for a more complete description.}
\label{fig:v_sigma_graph}
\end{figure}

Finally, we wish to comment on a simple tool commonly used to assess the importance of rotation in determining the global shape of a stellar system, that is the plot $V/\sigma$ vs. $\varepsilon$ (in the context of elliptical galaxies, see \citealp{Davies1983} and \citealp{Emsellem2011}). In Fig.~\ref{fig:v_sigma_graph}, the quantity $V/\sigma$ is the ratio of the observed maximum of the line-of-sight rotation profile to the central line-of-sight velocity dispersion, as reported in Col.~(3) of Table~\ref{tab:3}; the values of the ellipticity $\varepsilon$ are those reported by WS87. We indicate the ($V/\sigma$,\,$\varepsilon$) pairs by filled symbols. The empty symbols show the effect of correcting these values for inclination (the assumed inclinations are those reported in Table \ref{tab:1}), as if the globular clusters were viewed ``edge-on'' ( $i=90^\circ$), following the procedure described in \citet{Cappellari2007}. In the figure, the dashed line indicates the relation expected for isotropic oblate rigid rotators viewed ``edge-on'', whereas the thin dotted lines indicate oblate rotators (viewed ``edge on") with different global anisotropy parameter $\delta$ \citep{Binney2005,Cappellari2007}. Even though rotation and pressure anisotropy vary significantly with radius,\footnote{In fact it has been proposed to make a diagram for a suitably defined luminosity-weighted average of $V/\sigma$; note that within the analytical family of models that we are using, this and other differential indicators of the role of rotation can be constructed in a straightforward way.} according to this diagram, the flattening observed in the three globular clusters could be argued to be originated by the presence of internal rotation. The deviations from the line of isotropic rotators may be interpreted as due to the combined effects of inclination, differential rotation, and pressure anisotropy. The physically simple self-consistent models that we have tested in this paper give insight into how the various physical ingredients may combine their roles into the observed data. The cluster for which the deviation is most significant is $\omega$ Cen, confirming its anomalous behavior (which we have argued to be the result of its only partially relaxed state). This result is even more striking if we refer to the points corrected for inclination.

\section{Summary and conclusions}
\label{conclusions}

In the present paper we have applied a family of self-consistent global dynamical models, recently constructed with the purpose of describing differentially rotating star clusters in a quasi-relaxed state, to three Galactic GCs, namely $\omega$~Cen, 47~Tuc, and M15, that exhibit evidence for flattening and rotation. For these clusters an extremely rich set of data is available, particularly on their three-dimensional kinematics. 

With respect to the traditional modeling of globular clusters, generally limited to a study of the surface brightness profile (but see the effort made in ZBV12), we have given here highest priority to the interpretation of the available kinematical data. This is a particularly challenging test for the models. In turn, the success of the models for the two clusters known to be in a sufficiently well relaxed state allows us to measure their internal structural parameters accurately and reliably, well beyond the reach of simpler and more idealized models.

The modeling procedure is based on three steps. (1) We identify the relevant range of the model parameters from the characteristics of the observed differential rotation. (2) We set the relevant physical scales by means of a standard fitting procedure on the photometric profile and the line-of-sight kinematic profiles. (3) We use the models thus fully identified to make definite, quantitative predictions on several other observational data not used in the first two steps, namely the anisotropy profile $\sigma_t/\sigma_R$, the proper-motion mean-velocity profiles $V_t$ and $V_R$, the ellipticity profile $\varepsilon$, and a map of the relevant projected isodensity contours; the quality and reliability of the adopted family of models is best assessed in this last predictive step, where we do not have free model parameters anymore available. Finally, by combining the gathered information on the proper-motion velocity-dispersion with that on the line-of-sight velocity profiles, we obtain a dynamical estimate of the distance to the cluster.

An application of nonspherical models to the full set of data available for these clusters, including proper motions, has been made only by \citet{vandeVen2006} for $\omega$ Cen and by \cite{vandenBosch2006} for M15, based on a Schwarzschild orbit-based modeling procedure. 
Both modeling techniques assume axisymmetry and allow for internal rotation and anisotropy in velocity space. Our physically simple models are based on a distribution function defined in terms of two integrals of the motion and applied under the hypothesis of a constant mass-to-light ratio, whereas the orbit-based models include the possibility of a varying mass-to-light ratio and the presence of a third integral of the motion. Therefore, the latter approach allows for a more general investigation with a relatively free and more complex structure of the pressure anisotropy profile. Despite these differences, we obtain consistent results for the total mass estimates, for the global mass-to-light ratios, and in particular for the presence of tangential anisotropy in the outer parts of $\omega$ Cen. To some extent, the descriptive orbit-based modeling and the predictive distribution-function based modeling are complementary: the fact that they lead to similar results is highly nontrivial and strengthens the conclusions that are obtained.

An important merit of using a physically based family of models is to make predictions. Indeed, we offer a hopefully general physical interpretation of the observed features (in particular, in relation to the interplay between rotation and anisotropy in determining the internal structure of the stellar systems) that might be tested soon on other clusters. In addition, for the three clusters considered in this paper, we make specific predictions about the structural properties in their outer parts which may be tested by future observations (e.g., anisotropy profile, rotation profile, isodensity contours).

The main results obtained in this paper are the following:

\begin{itemize}

\item For the three most studied globular clusters we have illustrated how such detailed modeling procedure can be implemented to make a test on the adequacy of a physically justified, global, self-consistent family of models to interpret all the available photometric and kinematic data, including a rich set of proper motions. One important technical problem, the inclination and projection of self-consistent models constructed from a nontrivial distribution function, has been resolved by the use of suitable discrete realization in terms of a large number of simulated particles (see Sect.~3.1).

\item For the well-relaxed cluster 47 Tuc the model that we have identified provides a very good interpretation of the photometric and kinematic data. In particular, the rotation profile is well matched throughout the entire extent of the cluster, showing clearly the position of the maximum rotation velocity, the characteristic rigid rotation behavior in the central region, and the relatively sharp decline in the outer parts. In addition, the proper-motion rotation measured by \citet{Anderson2003} is well consistent with the value predicted by our model at the relevant radial positions. The identification of the model comes out naturally and leads to a specific prediction on the ellipticity profile that is in excellent agreement with the observations. 

\item For the relaxed cluster M15 we provide a global model in good agreement with the data; in particular, the line-of-sight rotation profile in the central regions, characterized by a solid-body behavior, is well accounted for by the model. The possible presence of a fast-rotating core on the small radial scale (where the observed photometric profile shows evidence of a post-core-collapse phase) does not appear to influence the quality of our global description.  

\item The model selected for $\omega$ Cen is unable to reproduce the steep central gradient in the line-of-sight velocity dispersion profile; in addition, the predicted ellipticity profile is somewhat offset with respect to the observed profile. We showed how these features are likely to reflect  the condition of only partial relaxation of the cluster, as confirmed by the measured high radial anisotropy. Still, somewhat surprisingly, our model provides a satisfactory global interpretation of the complex three-dimensional kinematics available for this object. In particular, the overall behavior of the anisotropy profile is successfully described, including the presence of tangential anisotropy in the outer parts of the system.

\item The results of this study confirm that indeed internal rotation is responsible for the observed flattening for at least two of the three clusters (47 Tuc and M15). For $\omega$~Cen there is no doubt that rotation is important; still, the discrepancy between predicted and observed ellipticity profile suggests that pressure anisotropy cooperates in determining its observed morphology. 

\item We have determined new dynamical estimates of the distances to the stellar systems under consideration. Before, this kind of analysis has been performed in only a few cases (in particular, see \citealp{vandeVen2006} and \citealp{Anderson2010}). It appears that the distance estimates based on dynamical models are generally lower compared to those derived from photometric methods, such as the analyses of variable stars (e.g., see \citealp{Bono2008}), and from other stellar indicators.  
 
\end{itemize}
 
Further interesting insights may come from the study of rotation in different environments, such as in low-mass stellar systems in the Magellanic Clouds, where GCs are known to be younger and flatter than the Galactic clusters. In particular, strong differential rotation may be a critical ingredient in determining the structure and internal dynamics of the class of the so-called ``ring clusters'' (see \citealp{HillZaritsky2006,Werchan2011}), which are characterized by an off-centered peak density profile. The presence of internal rotation may play an important role also in the dynamics of low-mass stellar systems, in the transition region between classical star clusters and dwarf galaxies (e.g., see the recent spectroscopic study of the rotating ultra-compact dwarf performed by \citealp{Frank2011}). 

We showed that rotation plays an important role in determining the structure of the three clusters considered in this paper, but that morphological information (as exemplified by the ellipticity profile of $\omega$ Cen) can be decisive in assessing the quality of a model. It remains to be ascertained how frequently is rotation the key dynamical factor and which GCs owe their shape instead mainly to external tides or simply to pressure anisotropy. New observational efforts to study the morphology of low-mass stellar systems (in particular, devoted to the measurement of ellipticity profiles, isophotal contours, and quadrupole moments) are thus highly desired. 

The inclination angle of the objects (assumed here to be axisymmetric) is a key ingredient in modeling the data. This quantity is difficult to measure. Here for the three clusters we have adopted the inclination values reported in the literature. Based on the experience developed in this paper, we have devised a new method to determine simultaneously distance and inclination angle for a given axisymmetric stellar system, by means of the combined use of proper motions and line-of-sight velocities under the only assumption that the underlying distribution function depends on the two classical integrals of the motion $f = f(E,J_z)$. We plan to present this result soon, in a separate paper.   

Finally, we wish to reiterate (see also ZBV12) that many key dynamical issues (such as a reliable estimate of the dark matter content, the search of dynamical signatures of a possible central IMBH, and the evaluation of the effects of mass segregation) can be addressed exclusively by considering appropriate kinematical data in detail (for a recent study with a generally similar approach, but limited to the study of line-of-sight kinematic data, see \citealp{Sollima2012}). We thus hope that the detailed study presented in this paper may mark the beginning of fruitful developments in the study of the dynamics of globular clusters and other small-mass stellar systems, beyond the application of exceedingly idealized spherical, nonrotating models so far used almost universally.

\acknowledgments
We are grateful to A. Sollima for providing us with the kinematic profiles of his best-fit Wilson rotating model for $\omega$ Cen. We would like to thank M. Bellazzini, M. Gieles, D.C. Heggie, G. van de Ven, F. van Leeuwen, and E. Vesperini for useful comments and conversations. Finally we wish to thank the Referee for many constructive remarks that have helped improve the quality of the paper. This work was partially supported by the Italian MIUR.

\bibliographystyle{aa} 
\bibliography{biblio} 

\appendix
\section{The construction of the kinematic data sets used in the paper}
\label{App.A}

For $\omega$ Cen, two different data sets of line-of-sight velocities are considered: 1589 line-of-sight velocities from \citet{Reijns2006} and 649 line-of-sight velocities from \citet{Pancino2007} for the central part of the cluster. After identifying the stars in common between the two samples, the one with the lower associated error is kept. The final combined data set is composed of $N_\mathrm{los}$=1868 data, reaching a radial extent of approximately half truncation radius, with an average error of 1.98 km s$^{-1}$. The proper motions data available are the ones from \citet{vanLeeuwen2000}, with a total of 9847 ground-based measurements, and the ones from \citet{Anderson2010}, with a total of 72\,970 HST measurements.  We treat the two data samples as distinct.
In the \citet{vanLeeuwen2000} data set each star is provided with a membership probability and is classified according to the disturbance of the image due to neighboring stars on a scale from 0 to 4 (i.e., from non-disturbed to highly disturbed stars). We decided to select a subsample composed of stars with a membership probability higher than 68\%, belonging to class 0, and with error measurements lower than 0.25 mas yr$^{-1}$ (for a similar selection, see \citealp{vandeVen2006}). We obtain a sample composed of $N_\mathrm{pm}$ = 2740 proper motions, with a radial extent of approximately half truncation radius and an average error of 0.16  mas yr$^{-1}$ (corresponding to 3.89 km s$^{-1}$ for an assumed distance of 5.2 kpc). The data set from \citet{Anderson2010} is composed of two fields: a central field within $R\la R_\mathrm{c}$ and a field positioned along the major axis between $0.7R_\mathrm{c}\la R\la2.5R_\mathrm{c}$. The average error of the data is 0.078~mas~yr$^{-1}$ (corresponding to 1.92 km s$^{-1}$ for an assumed distance of 5.2 kpc).

For 47 Tuc, the line-of-sight velocities data set results from two data sets combined by following the procedure described in ZBV12: 499 line-of-sight velocities from \citet{Gebhardt1995} for the inner region ($R<100$ arcsec) and 1977 line-of-sight velocities from \citet{Lane2011} for the outer parts ($R>100$ arcsec). As noted in \citet{Lane2011}, the latter data set shows a mean velocity of $-16.85$~km~s$^{-1}$, which differs significantly from the value obtained from the former data set, $-18.34$ km s$^{-1}$; this is likely to be due to a systematic uncertainty between the zero-point of the two velocity systems. To correct for this offset we have subtracted from each data set the corresponding measured mean velocities. The final line-of-sight velocities sample is composed of $N_\mathrm{los}$ = 2476 velocities covering the entire extent of the cluster and with an average error of 2.29 km s$^{-1}$. The proper motions are taken from \citet{McLaughlin2006}, which is composed of N$_{pm}$ = 12\,974 HST proper motions selected on the basis of the star magnitude (V$<$20) and quality (i.e., we consider data with probability $P(\,\chi^2)>0.001$); unfortunately, the data cover only the central region out to $\approx$100 arcsec (approximately 4 core radii); the measurements have an average error of 0.27 mas yr$^{-1}$ (corresponding to 5.76 km s$^{-1}$ at a distance of 4.5 kpc).

For M15 we used a single data set composed of $N_\mathrm{los}$ = 1777 line-of-sight velocities from \citet{Gebhardt2000}; this sample is centrally concentrated, with $\approx 80$\% of the stars being inside $10\,R_c$ and with an average error of 3.79 km s$^{-1}$. In addition, we used the sample of $N_\mathrm{pm}$ = 703 HST proper motions in the central region of the cluster (R$<2\,R_c$), as reported by \citet{McNamara2003}, with an average error of 0.14 mas yr$^{-1}$ (corresponding to 6.79 km s$^{-1}$ at a distance of 10.2 kpc).

We recall that the procedure used to obtain the proper motions data sets described above will not reveal any solid body rotation in the plane of the sky, as well as any systematic motions of contraction or expansion (e.g., see \citealp{Vasilevskis1979,McLaughlin2006,Anderson2010}), because the proper motions measurements are relative measurements  (no absolute reference frame is available for measuring the star displacements at different epochs). \citet{vandeVen2006} show how to compensate for the missed solid body component under the assumption of axisymmetry in the proper motions sample of \citet{vanLeeuwen2000}, by combining line-of-sight velocities and proper motions. We apply the related correction to the $\omega$ Cen proper motions sample of  \citet{vanLeeuwen2000}, while we do not correct the one from \citet{Anderson2010}. For 47 Tuc and M15, given the fact that the data sets are centrally concentrated, we argue that, in the very central regions of the clusters, the amount of solid body rotation associated with this effect is negligible and therefore we do not apply any correction (see \citealp{vandenBosch2006}, who first noted that the result of the correction for M15 is below the measurement errors and therefore can be ignored). Therefore, for the last two clusters no sign of rotation in the plane of the sky is expected from the proper motions data sets considered above; however, rotation in the plane of the sky has been clearly detected for 47 Tuc by \citet{Anderson2003}, using as an absolute reference the background stars of the Small Magellanic Cloud.

Finally, an additional correction is applied to the $\omega$ Cen and 47 Tuc data, to correct for the apparent rotation resulting from their large angular extent and their global orbital motion in the Galaxy; to this purpose, we followed closely the procedure described by \citet{vandeVen2006}.

\section{Tests on the determination of the rotation position angle and amplitude}
\label{App.B}

To checked whether the rotation patterns of the GCs under study show radial variation of the position angle and of the rotation amplitude, we repeated the procedure outlined in Sect. \ref{rotation}, on subsamples of data with $R<R_\mathrm{max}$, for decreasing values of $R_\mathrm{max}$. Table \ref{tab:2} lists the results of the position angles and rotation amplitudes for given values of $R_\mathrm{max}$. 
To assess whether the number of data available for the different cases is sufficiently large to reach a significant measure of the position angles and of the rotation amplitudes, we tested the method used on simulated data drawn from a rotating model of the family introduced in Sect.~\ref{selection}. We found that the estimates of the position angles obtained from samples of data with $N\la100$ have a typical uncertainty (associated with a 68\% confidence level) greater than $\pm25\degr$. We conclude that no significant PA variation is present in 47 Tuc and $\omega$ Cen, whereas for M15 a twisting is detected from $260\degr$ in the innermost region (on the scale of the core radius) to $106\degr$ in the outer parts (thus confirming the result found by \citealp{Gebhardt2000}).

Moreover, we found that the rotation amplitude $A$ changes across the clusters. In general, it reaches a maximum at intermediate values of $R_\mathrm{max}$. This can be taken as an indication of differential rotation (as illustrated by the shape of the rotation profiles, see Figs.~\ref{fig:fit_omega}, \ref{fig:fit_47tuc}, and \ref{fig:fit_M15}). Interestingly, all three clusters show a sharp increase of the rotation amplitude in the very central regions. This feature may be interpreted as a signature of a complex rotation pattern, characterized by a rapidly spinning core, as reported by \citet{vandeVen2006} and \citet{vandenBosch2006}, ascribed to a disk-like rotating component in $\omega$ Cen and a decoupled rotating core in M15. The last rows in Table~\ref{tab:2} show that $\omega$ Cen reaches an amplitude of $A=13.93$ km~s$^{-1}$ for $R<0.5\,R_c$, 47 Tuc $A=4.78$ km~s$^{-1}$ for $R<0.6\,R_c$, and M15 $A=13.00$~km~s$^{-1}$ for $R<0.4\,R_c$.

To test the significance of the detected central rotation we performed a Monte Carlo simulation. We draw from a nonrotating model, characterized by a realistic value of the central concentration, a simulated data set with an equal number of data and similar spatial distribution with respect to the real case (see last row of Table \ref{tab:2}). We then computed for each cluster N=1000 random realizations of this synthetic data set and we applied to them the procedure to calculate the rotation amplitude $A$, as described above in Sect. ~\ref{rotation}. Finally, from the distribution of the derived rotation amplitudes, we calculated the probability of finding a value of $A$ higher than the one derived from the real data. We found that the probability of measuring by chance rotation amplitudes as high as the ones determined when no rotation is present is 7\%, 32\%, and $\la1$\% for $\omega$ Cen, 47 Tuc, and M15, respectively. We thus conclude that the central increase measured in 47 Tuc is not statistically significant, whereas it can be taken as a sign of genuine high rotation in the central regions of M15; this interpretation marginally applies also to the case of $\omega$ Cen.

\begin{table*}[h]
\caption{Internal rotation: position angle of the rotation axis and rotation amplitude referred to disks of different radii.}
\label{tab:2}
\centering
\begin{tabular}{lccc}
\hline\hline
\multicolumn{4}{c}{$\omega$ Cen}\\
$R_\mathrm{max}$&$A$&PA& $N$\\
(1)&(2)&(3)&(4)\\
\hline
\textit{all}& 6.79 & 12 & 1868\\
$10\, R_c$& 6.91 & 12 & 1827\\
$8\, R_c$& 7.09 & 10 & 1737\\
$6\, R_c$& 7.73 & 7 & 1481\\
$4\, R_c$& 7.58 & 11 & 1026\\
$2\, R_c$& 6.95 & 22 & 398\\
$1\, R_c$& 3.97 & 57 & 91\\
$0.9\, R_c$& 1.25 & $-4$ & 73\\
$0.7\, R_c$& 1.98 & 23 & 42\\
$0.6\, R_c$& 7.04 & 10 & 27\\
$0.5\, R_c$& 13.93 & $-8$ & 19\\
\hline
\end{tabular}\quad\quad\quad\quad
\begin{tabular}{lccc}
\hline\hline
\multicolumn{4}{c}{47 Tuc}\\
$R_\mathrm{max}$&$A$&PA&$N$\\
(1)&(2)&(3)&(4)\\
\hline
\textit{all}& 4.00 & 136 & 2476\\
$80\, R_c$& 4.11 & 136 & 2414\\
$40\, R_c$& 4.41 & 137 & 2058\\
$20\, R_c$& 4.53 & 136 & 1358\\
$10\, R_c$& 3.32 & 139 & 800\\
$5\, R_c$& 2.24 & 164 & 526\\
$2\, R_c$& 2.64 & 180 & 388\\
$1\, R_c$& 4.07 & 199 & 114\\
$0.8\, R_c$& 4.05 & 171 & 78\\
$0.7\, R_c$& 5.99 & 167 & 61\\
$0.6\, R_c$& 4.78 & 206 & 39\\
\hline
\end{tabular}\quad\quad\quad\quad
\begin{tabular}{lccc}
\hline\hline
\multicolumn{4}{c}{M15}\\
$R_\mathrm{max}$&$A$&PA&$N$\\
(1)&(2)&(3)&(4)\\
\hline
\textit{all}& 2.84 & 106 & 1777\\
$30\, R_c$& 2.89 & 106 & 1671\\
$10\, R_c$& 2.93 & 102 & 1467\\
$8\, R_c$& 3.00 & 99 & 1293\\
$5\, R_c$& 1.94 & 118 & 916\\
$4\, R_c$& 1.43 & 140 & 724\\
$2\, R_c$& 2.14 & 147 & 319\\
$1\, R_c$& 1.19 & 253 & 128\\
$0.6\, R_c$& 4.68 & 272 & 65\\
$0.5\, R_c$& 6.95 & 253 & 52\\
$0.4\, R_c$& 13.00 & 261 & 31\\

\hline
\end{tabular}
\tablecomments{For each cluster we report the value of the position angle of the rotation axis PA measured in degrees East of North [Col. (3)] and the rotation amplitude A in km s$^{-1}$ [Col.~(2)] obtained from a fit of a sine function when considering $N$  data [Col.~(4)] inside $R_\mathrm{max}$ [Col.~(1)]. For each cluster, the first row corresponds to the results illustrated in Fig.~1.}
\end{table*}
\end{document}